\documentclass[final,3p,times,authoryear]{elsarticle}

\usepackage{amssymb}

\usepackage{lineno}

\usepackage{float}
\usepackage{amsmath}
\usepackage{bm}
\usepackage{booktabs}
\usepackage{multirow}

\usepackage[utf8]{inputenc}
\usepackage[OT4,T1]{fontenc}
\usepackage{lscape, graphicx}
\usepackage{pdfpages}
\usepackage{array}
\usepackage{amsfonts,amsthm} 
\usepackage{subfig}
\usepackage{url}
\usepackage{rotating}
\usepackage{appendix}
\usepackage{xcolor}
\usepackage{epstopdf}

\journal{Ecological Indicators}

\begin{document}

\begin{frontmatter}



\title{A Strong Sustainability Paradigm Based Analytical Hierarchy Process (SSP-AHP) Method to Evaluate Sustainable Healthcare Systems}






\author[1,2]{Jaros\l{}aw W\k{a}tr\'{o}bski\corref{mycorrespondingauthor}}

\author[1,3]{Aleksandra B\k{a}czkiewicz}
\fntext[e-mail]{aleksandra.baczkiewicz@phd.usz.edu.pl (A. B\k{a}czkiewicz)}

\author[4]{Iga Rudawska}
\fntext[e-mail]{iga.rudawska@usz.edu.pl (I. Rudawska)}

\address[1]{Institute of Management, University of Szczecin, ul. Cukrowa 8, 71-004 Szczecin, Poland}
\address[2]{National Institute of Telecommunications, ul. Szachowa 1, 04-894 Warsaw, Poland}
\address[3]{Doctoral School of University of Szczecin, ul. Mickiewicza 16, 70-383 Szczecin, Poland}
\address[4]{Institute of Economics and Finance, University of Szczecin, ul. Mickiewicza 64, 71-101 Szczecin, Poland}

\cortext[mycorrespondingauthor]{jaroslaw.watrobski@usz.edu.pl (J. W\k{a}tr\'{o}bski)}

\begin{abstract}
The recent studies signify the growing concern of researchers towards monitoring and measuring sustainability performance at various levels and in many fields, including healthcare. However, there is no agreed approach to assessing the sustainability of health systems. Moreover, social indicators are less developed and less succinct. Therefore, with this work, authors seek to map sustainable reference values in healthcare and propose a conceptual and structured framework that can guide the measurement of the social sustainability-oriented health systems. Based on a new multi-criteria method called Strong Sustainability Paradigm based Analytical Hierarchy Process (SSP-AHP), the presented approach opens the availability for systems’ comparison and benchmarking. The Strong Sustainability Paradigm incorporated into the multi-criteria evaluation method prevents the exchangeability of criteria by promoting alternatives that achieve good performance values on all criteria, implying sustainability. The research results offer insights into the core domains, sub-domains, and indicators supporting a more comprehensive assessment of the social sustainability of health systems. The framework constructed in this study consists of five major areas: equity, quality, responsiveness, financial coverage, and adaptability. The proposed set of indicators can also serve as a reference instrument, providing transparency about core aspects of performance to be measured and reported, as well as supporting policy-makers in decisions regarding sectoral strategies in healthcare. Moreover, the presented approach can be helpful while performing cross-national comparisons. Our findings suggest that the most socially sustainable systems are Nordic countries. They offer a high level of social and financial protection, achieving very good health outcomes. Furthermore, their social sustainability is safeguarded by balancing all identified values. On the other hand, the most unsustainable systems, located in central and eastern European countries, have a high level of unmet needs, experiencing shortages in the workforce and underfunding of healthcare. A key challenge for them is to achieve sustainability by increasing the share of public expenditure on health and limiting the substantial dependence on out-of-pocket payments.
\end{abstract}




\end{frontmatter}


\section{Introduction}
\label{sec:intro}
In the new millennium sustainability is top on the agenda of international bodies and agencies such as United Nations, European Commission and Organization for Economic Cooperation and Development (OECD). Owning to Agenda 2030~\citep{un2015} and its 17 goals, the whole systems, industries and organizations have been called to rethink, manage and reduce the impact they have on economies, people and the planet.

The concept of sustainability has been well established in manufacturing industries~\citep{ali2021sustainable, gouda2020pressure} and is gaining more and more recognition in service sector~\citep{hussain2016framework}. Its understanding is not limited to environmental issues (e.g. waste management and pollution) and economic issues (e.g. cost-efficiency), but covers a wide spectrum of problems concerning welfare and wellbeing of various stakeholders. Nevertheless, relatively extensive concern has been expressed so far towards economic ~\citep{seuring2008literature, visser2002beyond} and environmental  ~\citep{saad2003integrated, sarkis2010stakeholder}goals of industries’ development. The importance of social aspect of sustainability increases with the complexity of systems and the multiplicity of stakeholders who are involved~\citep{ajmal2018conceptualizing, colantonio2009social}. The inclination towards obtaining social sustainability in the long run is reinforced by political agendas, global transformations~\citep{murphy2012social}, and humanitarian crisis such as COVID pandemic. It refers predominantly to healthcare sector, that is responsible for providing services to promote, restore, and improve the health of population~\citep{murray2000framework}. Moreover, apart from health improvement, actions undertaken within health systems should be responsive  and fair in financial contribution~\citep{popescu2018investigating}. These intrinsic goals formulated towards health systems seem to correlate with the people-oriented dimension of social pillar of sustainability. It emphasizes the wellbeing of individuals and the fair distribution of resources and that is why it seems to be the most relevant for the environment of healthcare~\citep{maghsoudi2020role}. 
 
To obtain and maintain social sustainability in the healthcare context, a health system ought to provide sufficient resources (human, financial, and physical), and processes to meet individual and public health and wellbeing. In spite of the growing interests expressed by many researchers towards the concept of social pillar of sustainability~\citep{hajirasouli2016social, huq2014social}, also in the context of healthcare~\citep{capolongo2016social, macassa2020rethinking, maghsoudi2020role}, there is still a little research on the systemic, macro-level factors contributing to the development of sustainability-oriented heath systems, highlighting the role of the social dimension. Moreover, there is no consensus so far how to monitor and measure the health system performance in terms of the sustainability. Most of the previous research focus on the organizational factors and explore the various motivators across healthcare supply chain~\citep{hussain2019exploration} and within the individual care facilities~\citep{aljaberi2020framework, cristiano2021systemic, jahani2021sustainability}. All of them take a micro perspective and  abstract from the sector-wide level. 

The literature review clearly demonstrates that MCDA methods have proven their usefulness in the area of sustainable healthcare systems assessment~\citep{puvska2021evaluation, ozturk2020new, zamora2021reconciling}. Furthermore, the analysis of the works mentioned above shows a great practical potential of MCDA methods from the so-called American school~\citep{dehe2015development}, including the AHP method~\citep{rahman2021multi}. The reasons for this are undoubtedly the simplicity of the algorithms themselves~\citep{aktas2015new} and easy adaptation of the AHP method to any form of the real evaluation problem~\citep{aljaberi2017framework, singh2014fuzzy}. On the other hand, it is possible to point out a significant disadvantage of the AHP method: the linear compensation criteria effect. The problem indicated in the practice term means that this method supports only the weak sustainability paradigm ~\citep{ziemba2017using}. It is particularly important in the practical development of sustainability assessment~\citep{ziemba2019towards} based models, where in practice (depending on the preference of the decision-maker or the purpose of the assessment), there is a need to model both strong and weak sustainability paradigms~\citep{ziemba2020multi}.

Therefore, the purpose of this work is to: a) map the sustainable values that would fit the context of the health system in terms of its social goals, b)  introduce a conceptual and structured framework that can guide the evaluation and measurement of the social sustainability-oriented health systems, c) develop of a multi-criteria method for evaluating healthcare systems oriented towards social sustainability, considering the strong sustainability paradigm. 

In the methodological term, an additional challenge undertaken by the authors is to take into account the dynamic influence of the decision-maker on the degree of criteria compensation, which in practice makes the method independent of the given evaluation problem and allows to model the level of strong sustainability paradigm arbitrarily. In order to prove the reliability of the proposed SSP-AHP method, its performance for expert weights determined is compared using the functions of the AHP method and Entropy and CRITIC. The functions included in the implemented AHP method, such as the eigenvector method, the normalized column sum method, and the geometric mean method, can be used both to determine the criteria weights based on decision-makers subjective indications and for multi-criteria evaluation of the sustainability of alternatives as well as a modification of the SSP-AHP with the strong sustainability paradigm~\citep{duleba2022introduction}. 

Rest of the paper is organised as follows. Section~\ref{sec:litrev} provides literature review. In section~\ref{sec:ConceptualFramework} the conceptual framework with indicators relevant in sustainable healthcare is detailed. Section~\ref{sec:methodology} introduces research methodology and framework. Obtained results are presented in section~\ref{sec:results} and discussed in~\ref{sec:discussion}. Finally, conclusions and future work directions are drawn in section~\ref{sec:conclusions}.

\section{Literature review}
\label{sec:litrev}
\subsection{Sustainability and social sustainability in healthcare context}

There are many definitions for sustainability in the literature. The most general one was proposed by World Commission on Environment and Development in 1987 and perceives sustainability as ''meeting the needs of the present without compromising the ability of future generations to meet their own needs''~\citep{wced1987world}. This interpretation lacks any criteria that could be indicative of the sustainability phenomenon. Since 1987 many other interpretations have been developed~\citep{sneddon2006sustainable}, that can be found suitable also for healthcare industry. However, there is currently no consensus in the field on which definition of sustainability to operationalise~\citep{lennox2020making}.  Sustainability can be considered at various levels. In narrow sense, it may refer to a program, clinical intervention, or strategy that continue to be delivered after a defined period of time and keep to produce benefits for the target group~\citep{moore2017developing}. Many research findings demonstrate the application of sustainability practices, across a range of healthcare settings, including primary, secondary, tertiary and community healthcare~\citep{lennox2020making}. In a broader sense, health system sustainability refers to the maintenance and constant adaptation to a changing environmental, social and economic context, assuring the effective and responsible use of limited healthcare resources in order to keep and improve the health and well-being of the population and of each individual~\citep{maghsoudi2020role}. It is in line with the WHO's goals for health systems, namely: improving health and  health equity in ways that are responsive, financially fair, and make most efficient use of available resources~\citep{who2007everybody}~\citep{who2000world, world2010world}. 

Given the above mentioned objectives of health systems social sustainability seems to be the most paramount in healthcare context. It is traditionally understood as a human-oriented dimension, that emphasizes the wellbeing of individuals and the fair distribution of resources~\citep{vuong2017psychological}. According to literature, this concept encompasses social issues, such as justice, equity and access to basic needs~\citep{awan2018governing}. Maghsoudi et al. indicate that such interpretation is appropriate in healthcare context, as it embodies  both the improvement of the wellbeing of patients and the calls for justice in the distribution of resources~\citep{maghsoudi2020role}. Additionally, Capolongo et al.~\citep{capolongo2016social} highlight that social sustainability implies also professionals’ well-being, safety, security and satisfaction. Other authors~\citep{chiu200312} stress the importance of providing equal opportunities to access health and safety resources and addressing the expectations of stakeholders, mainly patients while assessing the social sustainability in healthcare. Nevertheless, the issue what measures to apply to sector-wide perspective and how to choose the relevant indicators remains unsolved.

Of course, there are other approaches that incorporate MCDA methods and analyze the dynamics of variation in outcomes over time~\citep{pianosi2016sensitivity}. An example is a work in which we considered data collected for five years for several major cities in the United States to temporally assess integrated health status using a composite index based on maximum entropy networks (MENets)~\citep{servadio2018optimal}. There are also approaches in multi-criteria sustainability assessment considering several indicators based on machine learning models~\citep{lawal2021orthogonalization} and stochastic optimization algorithms in multi-criteria analytical decision models~\citep{convertino2019information}.

\subsection{The assessment of sustainability in healthcare}

Sustainability assessment seeks to monitor and evaluate the extent to which various aspects of a given organization, industry or system meets key goals of sustainable development~\citep{bond2013challenges}. In that sense it is a kind of performance evaluation that is seen as a process of collecting, computing and communicating quantified constructs for the managerial purposes~\citep{kollberg2005design, lizarondo2014assisting}, at various levels: global, national, sector-wide and company-level. The goal of this process is always the same - to follow up, monitor and improve the performance of the given system or organization. Although, there is a growing evidence on how to assess corporate sustainability~\citep{hallstedt2017sustainability, huge2013discourse, moldavska2016development, moldavska2019holistic, morrison2015handbook, pope2015conceptual, pope2017reconceptualising, singh2014fuzzy, waas2014sustainability, ziout2013multi}, and sustainability of public services~\citep{bandeira2018fuzzy}, there is no agreed approach to measuring sustainability in healthcare context, especially at the sector-wide level. Few works in this field have focused on the organizational level and aimed at developing the assessment criteria that were linked to strategic planning of the health service organisation~\citep{hussain2016framework}. AlJaberi et al.~\citep{aljaberi2020framework} introduced an interesting assessment framework, based on analytical hierarchical process (AHP), proposing factors that cover core business functions of healthcare organization. The concept is very interesting, but does not correspond strictly to the sector-wide level. In turn, Noveiri and Kordrostami~\citep{jahani2021sustainability} proposed an approach, based on data envelopment analysis (DEA), applied to measure the sustainability performance of hospitals in the presence of imprecise measures. Although, the findings disclose informative details about overall sustainability performance in hospitals, they cannot be directly applied at the sector-wide level. The similar conclusion can be drawn from the framework introduced by Sarriot et al.~\citep{sarriot2004methodological}, that places sustainability at the center of primary health care programming. This promising methodology concerns the assessment of sustainability in NGO primary health care and community-based projects.

Although, social sustainability and its measurement are gaining more and more recognition in the literature dedicated to manufacturing~\citep{rajak2016sustainable} and services~\citep{kumar2019development}, and even cities and regions~\citep{dempsey2011social, yuan2020evaluating}, a few studies have contributed to delineating the significance of social sustainability in healthcare industry. For instance, Hussian et al.~\citep{hussain2019exploration} presented an AHP model to explore and assess the motivators of social sustainability in healthcare supply chain context. The work captures various motivating practices such as media and reputation, excellence and awards, organizational practice, technology and innovation, and attitudes. On the other hand, core categories like humanization, comfort and distribution were covered by Capolongo et. al.~\citep{capolongo2016social} to assess the social sustainability of hospital facility. In other studies, the social dimension has been designed for assessing customer, employee, and community well-being impacted mainly by manufacturing activities and manufactured products~\citep{eslami2019survey}. The range and the scope of the assessment of social pillar was wide and diverse, indicating the categories that reflected the attitude of a company towards its stakeholders and societal impact of the product (for example customer satisfaction, occupational health and safety, security and wages of employees, their satisfaction, education and training, product responsibility~\citep{eastlick2012increasing, huang2017sustainable, lu2011framework, rezvan2014sustainability}). All works cited demonstrate valued contribution to the understanding of sustainability at the organizational level, but their findings have poor adaptability to the context of a whole-sector assessment. Therefore, with this work authors seek to identify multiple criteria which are broad in scope and help to obtain comprehensive picture of social sustainability of health system.

\subsection{Multi-criteria methods in healthcare domain}

As it was presented in the previous section, different quantitative multi-criteria approaches are applied in healthcare assessment~\citep{aljaberi2017framework, puvska2021evaluation, torkayesh2021integrated}. In this section, MCDA-based approaches will be analyzed in detail in the terms of typology of models and methods used. An overview of MCDA methods used in the healthcare domain is summarized in Table~\ref{tab:litrevMCDAhealth} considering methods from American school, namely AHP~\citep{nemeth2019comparison}, MAVT~\citep{angelis2017multiple}, CRADIS~\citep{puvska2021evaluation}, fuzzy TOPSIS and VIKOR~\citep{ozturk2020new}, MACBETH~\citep{pereira2020using}, CoCoSo~\citep{torkayesh2021integrated}, DEMATEL~\citep{ortiz2016integrated}, MAUT~\citep{zamora2021reconciling}, and methods representing European school, such as ELECTRE TRI-NC~\citep{rocha2021quality} and PROMETHEE~\citep{makan2021sustainability}. Most of the works cited are based on methods from the American school. Among them, the AHP method is one of the most popular and widely used methods for sustainability assessment in the healthcare field, both for the location of medical sites and health systems assessment, especially when a hierarchical model with a large number of criteria is to be evaluated~\citep{dell2018combining}. Weights are most often determined by experts and decision-makers using methods such as AHP, SWING, BWM~\citep{muhlbacher2016making}. The advantage of the AHP method is an easy adaptation to hierarchical structure often found in models of sustainable healthcare problems easy use even for many alternatives under evaluation~\citep{dehe2015development}. The AHP method is also used in healthcare fields such as assessing sustainable service quality~\citep{aktas2015new, singh2019measuring}, assessing sustainable electronic services quality~\citep{buyukozkan2012combined}, general measurement of sustainability~\citep{aljaberi2017framework} and selecting sustainable healthcare infrastructure site location~\citep{dehe2015development}.

However, there are also papers presenting the application of methods from the European school. Among them is PROMETHEE, which application for assessment of the sustainable healthcare waste treatment system is presented in paper~\citep{makan2021sustainability}. PROMETHEE belongs to the outranking methods family, similar to ELECTRE family methods. Among the examples of applying methods belonging to the ELECTRE family in healthcare systems assessment, it is worth mentioning the application of the ELECTRE TRI-NC method for assessing hospital sustainability and quality~\citep{rocha2021quality}. A low level of criteria compensation characterizes both ELECTRE and PROMETHEE families of MCDA methods, so they are suitable for evaluating multi-criteria problems that focus on sustainability.
%
%
\begin{table}[H]
\centering
\caption{MCDA methods applied to sustainability healthcare assessment problem in reviewed literature.}
\label{tab:litrevMCDAhealth}
\resizebox{0.9\linewidth}{!}{
\begin{tabular}{
>{\raggedright\arraybackslash}p{5cm}
>{\raggedright\arraybackslash}p{6cm}
>{\raggedright\arraybackslash}p{8cm}
>{\centering\arraybackslash}p{3cm}
} \toprule
Authors & Problem & Methods applied & Number of Criteria/Sub-criteria \\ \midrule
\citep{dehe2015development} & Selection of sustainable healthcare infrastructure site location & ER (Evidential Reasoning) and AHP (Analytical Hierarchy Process) for determining criteria weights & 7/28 \\
\citep{aktas2015new} & Assessment of sustainable service quality in healthcare field & AHP & 7/28\\
\citep{singh2019measuring} & Assessing the quality of sustainable health services & Fuzzy AHP & 6/30 \\
\citep{aljaberi2017framework} & Measurement of sustainability in health systems & AHP & 7/31 \\
\citep{buyukozkan2012combined} & Assessment of sustainable electronic services quality in healthcare industry & Fuzzy AHP and fuzzy TOPSIS & 6/20 \\
\citep{angelis2017multiple} & Sustainable health technology assessment including new medicines evaluation & MAVT (Multi-Attribute Value Theory) &  4/14 (preliminaries based on literature review) and 5/11/28 (advanced practical assessment) \\
\citep{puvska2021evaluation} & Evaluation of sustainable healthcare waste & FUCOM (FUll COnsistency Method) for determining weights and CRADIS (Compromise Ranking
of Alternatives from Distance to Ideal Solution)
 & 4/16 \\
\citep{ozturk2020new} & Selecting the most sustainable health technology & AHP for determining weights, fuzzy TOPSIS (Technique for~Order of~Preference by Similarity to Ideal Solution), fuzzy VIKOR (VlseKriterijumska Optimizacija I Kompromisno Resenje), goal programming & 9/45/115/113 \\
\citep{pereira2020using} & Evaluation sustainability of European health systems & MACBETH (Measuring Attractiveness by a Categorical Based Evaluation Technique) & 3/6/9 \\
\citep{nemeth2019comparison} & Sustainable decision making in healthcare & SWING, SMART (Simple Multiattribute Rating Technique), AHP (Analytical Hierarchy Process), MACBETH, DCE (Discrete Choice Experiments), PAPRIKA (Potentially All Pairwise Rankings of All Possible Alternatives), CA (Conjoint Analysis) for criteria weighting & Not specified \\
\citep{torkayesh2021integrated} & Assessment of a sustainable health care sector in selected European countries & BWM (Best-Worst Method) and LBWA (Level Based Weight Assessment) for criteria weights determination, CoCoSo (Combined Compromise Solution) for evaluation of healthcare performances of several countries & 7 \\
\citep{ortiz2016integrated} & Selecting the most sustainable allied hospital variant for an integrated network of laboratory services & DEMATEL (Decision Making Trial and Evaluation Laboratory) for evaluation the independence between factors of the same category, AHP for calculating weights of criteria and categories & 4/14 \\
\citep{rocha2021quality} & Assessing hospital sustainability and quality & ELECTRE TRI-NC & 5/24 \\
\citep{zamora2021reconciling} & Assessing sustainable health profit in the context of financial attributes affecting consumption & MAUT (Multi-Attribute Theory) & 3/7/2 \\ 
\citep{makan2021sustainability} & Assessment of the sustainability of healthcare waste treatment systems & surrogate weighting and PROMETHEE (Preference Ranking
Organization METHod for Enrichment Evaluations) & 4/16 \\
\bottomrule
\end{tabular}
}
\end{table}

\subsection{Strong sustainability paradigm in MCDA methods}
\label{sec:strongParadigm}

An important aspect in a reliable sustainability assessment with MCDA methods is the proper modeling of criteria compensation~\citep{ziemba2017using}. In practice, there exist two paradigms: strong and weak sustainability~\citep{ziemba2017using}. The weak sustainability paradigm assumes criteria compensation, implying that individual evaluation criteria are interchangeable and can replace each other~\citep{norouzi2021comparison}. On the other hand, strong sustainability considers the complementarity of criteria but not interchangeability~\citep{ziemba2017using}. High compensation means that favorable performance values of several criteria can compensate for the poor performance of other criteria. Therefore, methods with high criteria compensation support a weak sustainability paradigm. This problem occurs to a higher or lower level in existing MCDA methods~\citep{ziemba2019towards}. Most MCDA methods are based on utility~\citep{rahman2021multi}, or value~\citep{stewart2016dealing} theory and these methods are usually based on a single synthesizing criterion. Among them, the popular AHP~\citep{seyedmohammadi2018application} method, TOPSIS, VIKOR, DEMATEL, SMARTER~\citep{ziemba2020multi} methods can be mentioned. However, these methods belonging to the so-called American school~\citep{wkatrobski2019generalised}, assuming full substitution of all criteria in the model, therefore only fulfill a weak sustainability paradigm, which for example, in the evaluation of healthcare is undesirable~\citep{oppio2018assessing}. The use of criteria weights partially reduces the criteria compensation problem, but this is insufficient to achieve the strong sustainability paradigm~\citep{ziemba2019towards}. Methods in which the compensation of criteria is limited include methods of the so-called European school~\citep{ezbakhe2018multi} using the outranking relation, for example, ELECTRE or PROMETHEE~\citep{makan2021sustainability}. However, methods from this group are characterized by a complex computational procedure~\citep{ziemba2019towards} and often recommend alternatives ranked in non-quantitative form~\citep{rocha2021quality}. Moreover, in the considered problem (healthcare sustainability assessment), often a large number of criteria is taken into account, forming a complex hierarchical model~\citep{dehe2015development}, which also significantly limits the possibility of model structuring~\citep{ortiz2016integrated} in the case of methods from the European school.

It is also worth noting that, from the point of view of the strong/weak sustainability paradigm, in methods from the European school, the degree of criteria compensation (although it is less than in the case of the AHP method) is still significant~\citep{ziemba2017using}. This fact encourages modifying the algorithms of the methods to reduce the degree of criteria compensation and thus introduce the paradigm of stronger sustainability to the known and widely used methods. For example, such an attempt was made in the case of the PROMETHEE II method, based on which the authors proposed the PROSA method~\citep{ziemba2017using} and its variations, namely generalized PROSA-G~\citep{ziemba2019towards}, PROSA-C~\citep{ziemba2020multi} to obtain a more sustainable evaluation of the variants~\citep{chmielarz2022assessment}. Successfully results of this attempt encourage to take analogous efforts, especially concerning methods with significant compensation of criteria, such as the widely used AHP method. Furthermore, the wide popularity of the AHP method in sustainable healthcare systems assessment motivated the authors to develop an extension of the AHP method supporting a strong sustainability paradigm called SSP-AHP, which is proposed in this paper.

\section{The conceptual framework - factors affecting social sustainability in health systems}
\label{sec:ConceptualFramework}

There are many criteria that could be used to assess the health system sustainability performance. Although, the recent studies signify the growing concern of researchers towards this issue~\citep{lennox2018navigating, lennox2020making}, there is no standard framework that would provide the universal set of criteria and indicators. The proposed cluster of criteria (see Table~\ref{tab:Criteria}) corresponds with the overall goals of health system defined by WHO and social sustainability objectives.

\begin{table}[H]
\centering
\caption{The summary of main criteria and sub-criteria in health system’s social-sustainability assessment.}
\label{tab:Criteria}
\resizebox{\linewidth}{!}{
\begin{tabular}{lp{3cm}p{7cm}lp{6cm}p{5cm}c}
\hline
Main criteria & \multicolumn{1}{l}{Sub-criteria} & \multicolumn{1}{l}{Relevant studies} & Criterion & Name & Unit & Goal \\ \hline
\multirow{7}{*}{$G_1$ - Equity} & \multirow{2}{3cm}{Service access} & \multirow{7}{7cm}{\citep{awan2018governing, capolongo2013measuring, capolongo2016social, hurst2001performance, levesque2020combining, murray2000framework}
} & $C_1$ & Cataract surgery performed in hospitals & [Number per 100\,000 inhabitants] & $\uparrow$ \\
 &  &  & $C_2$ & Hip replacement performed in hospitals & [Number per 100\,000 inhabitants] & $\uparrow$ \\
 & \multirow{2}{3cm}{Availability of human resorces} &  & $C_3$ & Practising physicians & [Density per 1\,000 population] & $\uparrow$ \\
 &  &  & $C_4$ & Practising nurses & [Density per 1\,000 population] & $\uparrow$ \\
 & \multirow{3}{3cm}{Availability of health infrastructure} &  & $C_5$ & Acute care beds & [Per 1\,000 population] & $\uparrow$ \\
 &  &  & $C_6$ & Computed Tomography scanners & [Per 1 million population] & $\uparrow$ \\
 &  &  & $C_7$ & Magnetic Resonance Imaging units & [Per 1 million population] & $\uparrow$ \\ \hline
\multirow{7}{*}{$G_2$ - Quality of care} & \multirow{4}{3cm}{Effectiveness of treatment} & \multirow{7}{7cm}{\citep{bankauskaite2007health, caunic2019frameworks, de2020oecd, giovanelli2015developing, hurst2001performance, kim2020developing, sarriot2004methodological, vuong2017psychological}} & $C_8$ & AMI 30 days standardised mortality & [Age-sex standardised rate per 100 patients] & $\downarrow$ \\
 &  &  & $C_9$ & Breast cancer five year net survival & [\%] & $\uparrow$ \\
 &  &  & $C_{10}$ & Lung cancer five year net survival & [\%] & $\uparrow$ \\
 &  &  & $C_{11}$ & Infant mortality & [Deaths per 1\,000 live births] & $\downarrow$ \\
 & Patient safety &  & $C_{12}$ & Observed percentage of hospitalized patients with at least one healthcare-associated infections & [\%] & $\downarrow$ \\
 & \multirow{2}{3cm}{Health outcomes} &  & $C_{13}$ & Life expectancy & [Years at birth] & $\uparrow$ \\
 &  &  & $C_{14}$ & Avoidable mortality & [Deaths per 100\,000 population] & $\downarrow$ \\ \hline 
\multirow{3}{*}{$G_{3}$ - Responsiveness} & Patients’ experience with economic burden & \multirow{3}{7cm}{\citep{araja2018opportunities, busse2013understanding, fujisawa2017measuring, levesque2020combining}} & $C_{15}$ & Self-reported unmet needs for health care due to financial reasons & [\%] & $\downarrow$ \\
 & \multirow{2}{3cm}{Patients’ experience  with non-economic burden} &  & $C_{16}$ & Self-reported unmet needs for health care due to distance or transportation & [\%] & $\downarrow$ \\
 &  &  & $C_{17}$ & Self-reported unmet needs for health care due to waiting list & [\%] & $\downarrow$ \\ \hline
\multirow{3}{*}{$G_{4}$ - Financial coverage} & Risk protection & \multirow{3}{7cm}{\citep{hurst2001performance, murray2000framework, popescu2018investigating}} & $C_{18}$ & Social health insurance (total healthcare) & [\%] of total population & $\uparrow$ \\
 & \multirow{2}{3cm}{Financial contribution} &  & $C_{19}$ & Public expenditures on health - share of current expenditure on health & [\%] & $\uparrow$ \\
 &  &  & $C_{20}$ & Public expenditure on health & [US Dollar per capita] & $\uparrow$ \\ \hline 
\multirow{5}{*}{$G_{5}$ - Adaptability} & \multirow{2}{3cm}{Investments in public health} & \multirow{5}{7cm}{\citep{demartini2017performance, lennox2018navigating, levesque2020combining, sarriot2004methodological, urquhart2020defining}
} & $C_{21}$ & Public expenditure on health - share of GDP & [\%] & $\uparrow$ \\
 &  &  & $C_{22}$ & Capital expenditure on health as a share of GDP & [\%] & $\uparrow$ \\
 & \multirow{2}{3cm}{Investments in human resources} &  & $C_{23}$ & Medical graduates & [Per 100\,000 population] & $\uparrow$ \\
 &  &  & $C_{24}$ & Nursing graduates & [Per 100\,000 population] & $\uparrow$ \\
 & New technologies uptake &  & $C_{25}$ & Proportion of primary care physician offices using electronic medical   records & [\%] & $\uparrow$ \\ \hline
\end{tabular}
}
\end{table}
The last ones are high on the political agenda of many international initiatives and health policies, focusing on strengthening health system accountability. For instance, Tallinn Charter signed by 53 states, promotes the Health 2020 policy framework to ensure sustainability of the system and the efficiency of healthcare expenditure~\citep{tello2015strengthening}.
The proposed framework is grounded in the literature on management by objectives~\citep{drucker2012practice} and goal setting~\citep{locke2013new} and perceives the definite goals as a core to assess performance. Table~\ref{tab:Criteria} presents the hierarchical structure of indicators of sustainable development of healthcare system. The model builds on the following dimensions which correspond to the social sustainability targets: 1) equity), 2) quality, 3) responsiveness, 4) financial coverage, and 5) adaptability. 
Moreover, we assume that the criteria included in the framework have to meet the WHO standards~\citep{world20182018}, namely: a) are prominent in monitoring of major international declarations (here : Agenda 2030, SG 3), b) are meeting the SMART formula c) there is a strong track record of extensive measurement experience, preferably supported by an international database (here: OECD Health Statistics and Eurostat), d)  they are used by countries in monitoring of national plans.

Universal access to health services has been formally accepted as the main goal of health systems~\citep{world2000equity}, encompassing both distribution of payments for health services and distribution of access to healthcare across population. Thus, equity can be considered as a cross-cutting dimension~\citep{barsanti2014equity}~\citep{levesque2020combining}. One of the most common metric that is widely accepted in WHO~\citep{murray2000framework} and OECD~\citep{hurst2001performance} frameworks to evaluate equity is access to healthcare.In our study we distinguish service access and availability of healthcare resources (both human and physical) as sub-criteria, which is in line with the findings by Capolongo et al.~\citep{capolongo2013measuring, capolongo2016social} and Awan et al.~\citep{awan2018governing}  It is here understood as assuring distribution of health resources and providing equal opportunities to access these resources. 

Quality is another most used domain of health systems’ performance assessment, widely recognized in the literature~\citep{bankauskaite2007health, giovanelli2015developing, kim2020developing}~\citep{hurst2001performance}~\citep{who2018}. In the revised OECD framework, released in 2006, quality of care is the core dimension and incorporates effectiveness, safety and responsiveness~\citep{caunic2019frameworks}. Quality indicators are often discussed in relation to health improvements, that is why we distinguished health outcomes as a sub-criterion in our model. Patient safety remains an important indicator of quality of care, that is high on health policy agenda. Risk mitigation, learning-based health systems, and healthcare environment design taking human factors into account remain in the main focus~\citep{fujisawa2017measuring}. Thus, enhancement of patient safety can be seem an essential component of social sustainability development in healthcare.

The growing concern towards patient’s centeredness, expressed by international bodies~\citep{fujisawa2017measuring}, allow us to distinguish system’s responsiveness as a separate criterion. It is here understood as ''the ability of the health system to meet the population’s legitimate expectations regarding their interaction with the health system, apart from expectations for improvements in health or wealth''~\citep{who2000}. The concept of responsiveness was introduced by WHO in 2000 in order to strengthen the social aspect of the health system and to underline the importance of human relationships within it~\citep{araja2018opportunities}. Responsiveness refers to ethical dimension of health system and covers the non-medical expectations of patients and their experience with the system~\citep{busse2013understanding, papanicolas2008principles}. Such features as respect, patient satisfaction, patient engagement, person centered care, quality of communication as well as patients’ insights into unmet needs are taken into consideration~\citep{levesque2020combining}. Assessment of responsiveness is possible through patients’ experience surveys, presenting their insights~\citep{matters2019patient}. Therefore, we proposed to measure responsiveness relying on two indicators derived from public opinion surveys across countries.

There is a general consensus, that healthcare should be provided without contributing to impacting financial strain and poverty for the individuals and population~\citep{popescu2018investigating}. Unfortunately, as reported by WHO~\citep{whoOrg} ''currently, at least half of the people in the world do not receive the health services they need. About 100 million people are pushed into extreme poverty each year because of out-of-pocket spending on health''. Thus, protecting people from the financial burden in scope of healthcare should be an important goal of social-sustainable health systems. Fairness in financial contribution plays a central role in WHO and OECD health system performance framework~\citep{murray2000framework}~\citep{hurst2001performance}, highlighting the importance of financial aspects. Our model encompasses both elements, defined as sub-criteria: risk protection and financial contribution. 

Last but not least, adaptability of a health system appears to be a crucial factor affecting its social sustainability. It is also an important element in performance assessment~\citep{demartini2017performance}. Adaptability is here understood as the system’s ability to flexible adjustment to the changing trends in healthcare market, across diverse contexts of delivery. As the health needs change, so the technologies available to meet them, systems need to be able to adapt~\citep{levesque2020combining}. Thus, pace of increase in public expenditure, investment in R\&D programs, and uptake of effective new technologies should be considered. Especially, the last issue seems to be the key to sustainability~\citep{lennox2018navigating, urquhart2020defining}. This also agrees with Sarriot et al.~\citep{sarriot2004methodological} who see management strategies that promote continual adaptability as a feature that guides system towards sustainable outcomes. 

\section{Research Methodology}
\label{sec:methodology}

This section presents the new multi-criteria method called Strong Sustainability Paradigm based Analytical Hierarchy Process (SSP-AHP) as well as the flow of the complete research framework with its particular steps in detail. In addition, data sources and methods for modeling criteria weights are also defined here.

\subsection{The SSP-AHP Method}

The proposed SSP-AHP method is based on the well-established and widely used AHP method, an aggregation method developed in 1980 by Saaty~\citep{dehe2015development}. Its advantages are its ease of use and scalability, which allows it to be easily applied even for complex hierarchical decision models. Moreover, the AHP method enables the determination of criteria weights besides evaluating alternatives. The usefulness of AHP arises from its usefulness in multi-criteria evaluation for both criteria weighting and evaluation of alternatives. For this reason, several online tools and software have been implemented and made open access, for example, AHP-OS~\citep{goepel2018implementation, chan2019prioritizing} and pyrepo-mcda~\citep{wkatrobski2022pyrepo, wkatrobski2022version}. The authors of this article have developed and made available a repository with the AHP method extended to model the compensation reduction of the SSP-AHP criteria for sustainability assessment~\citep{sspahpgithub2022}. The mentioned tools made available to a broad scientific audience significantly enhance analytical capabilities using the AHP method and contribute to its expansion with additional functionalities. On the other hand, the disadvantages of AHP include the phenomenon of criteria compensation presented in the section~\ref{sec:strongParadigm} and the possibility of reversal of ranking when an alternative is added or removed from the evaluated set. However, in the case of the sustainable healthcare evaluation problem, the set of alternatives, i.e., countries, is stable, so the rank reversal phenomenon need not be considered here. On the other hand, solving the problem of criteria compensation in AHP is the attempt to involve the SSP-AHP method development undertaken by the authors in this paper.

Because the proposed SSP-AHP method is the main research method for investigation presented in this paper, the formal assumptions and the algorithm of the proposed method are presented. The SSP-AHP method (the Strong Sustainability Paradigm based Analytical Hierarchy Process method) is an extension of the classical AHP method with the possibility of modeling the degree of compensation of individual criteria indicated by decision-maker, which is in accordance with the paradigm of strong sustainability. The main advantage of this method is recommending more sustainable decision variants and enabling investigating the sustainability of alternatives considering all criteria or their groups. This method complements the classical AHP with the measure of the mean absolute deviation $MAD$ of alternatives' performances extended by the value of the sustainability coefficient $s$. Coefficient $s$ represents the degree of reduction in criteria compensation. The subsequent steps of the SSP-AHP method are as follows.

\textbf{Step 1.} Normalization of the decision matrix. The SSP-AHP method, similarly to each MCDA method, requires a decision matrix demonstrated by Equation~(\ref{eq:decisionmatrix})

\begin{equation}
    X = [x_{ij}]_{m \times n} = \left [ \begin{array}{cccc} 
    x_{11} & x_{12} & \cdots & x_{1n} \\
    x_{21} & x_{22} & \cdots & x_{2n} \\
    \vdots & \vdots & \vdots & \vdots \\
    x_{m1} & x_{m2} & \cdots & x_{mn} \\
    \end{array} \right ]
    \label{eq:decisionmatrix}
\end{equation}
where $m$ denotes the number of alternatives, and $n$ represents the number of criteria. The SSP-AHP method proposed by the authors takes as input a matrix containing the quantitative performance values of the criteria, as the research addresses a problem in which such values are made available. The normalization procedure can be performed by applying any normalization method. This research employs the Minimum-Maximum normalization method, which is conducted using the Equation~(\ref{eq:minmaxProfit}) for profit criteria, while for cost criteria Equation~(\ref{eq:minmaxCost}) is applied.

\begin{equation}
    r_{ij} = \frac{x_{ij}-min_{j}(x_{ij})}{max_{j}(x_{ij})-min_{j}(x_{ij})} \label{eq:minmaxProfit}
\end{equation}

\begin{equation}
    r_{ij} = \frac{max_{j}(x_{ij})-x_{ij}}{max_{j}(x_{ij})-min_{j}(x_{ij})} \label{eq:minmaxCost}
\end{equation}

\textbf{Step 2.} The next step involves calculating $MAD_{ij}$ of each value in the matrix $R$, according to Equation~(\ref{eq:mad})

\begin{equation}
    MAD_{ij} = |\overline{r_{j}} - r_{ij}|s_{j} \label{eq:mad}
\end{equation}
where $R = [r_{ij}]_{m \times n}$ denotes normalized decision matrix, $m$ represents number of alternatives ($i = 1, 2, \ldots m$) and $n$ denotes number of criteria ($j = 1, 2, \ldots n$) and $s$ denotes sustainability coefficient. This coefficient can take values from 0\% implying no criteria compensation reduction at all, which is equivalent to classical AHP, to 100\%, denoting full criteria compensation reduction. A high $s$ coefficient indicates a high degree of criteria compensation reduction.

\textbf{Step 3.} Then from the value of the normalized decision matrix, $MAD_{ij}$ is subtracted, like Equation~(\ref{eq:matrixB}) demonstrates.

\begin{equation}
    b_{ij} = r_{ij} - MAD_{ij} \label{eq:matrixB}
\end{equation}

\textbf{Step 4.} This step involves an aggregation procedure involving calculating the utility function values for each alternative performing an aggregation using a weighted sum of $b_{ij}$ values, as Equation~\ref{eq:utility} shows.

\begin{equation}
    U_{i} = \sum_{j=1}^{n}b_{ij}w_{j} \label{eq:utility}
\end{equation}
where $w$ denotes weight value for each $j$th criterion ($j = 1, 2, \ldots, n$). Weights are determined using methods provided in~\ref{sec:appWeights}.

\textbf{Step 5.} The last step of the method is to order the alternatives in descending order according to the utility function values of the $i$th alternative obtained in the previous step. The alternative with the highest value of $U_{i}$ is the best option. 

\subsection{The Research Framework}

This subsection demonstrates the research framework and results provided by the SSP-AHP method. To evaluate the European countries listed in Table~\ref{tab:countries}, the authors used a hierarchical set of criteria. 

\begin{table}[H]
\centering
\caption{Countries selected for social sustainability-oriented health systems evaluation.}
\label{tab:countries}
\resizebox{0.7\linewidth}{!}{
\begin{tabular}{llllllll} \toprule
$A_{i}$ & Country & $A_{i}$ & Country & $A_{i}$ & Country & $A_{i}$ & Country \\ \midrule
$A_{1}$ & Belgium & $A_{5}$ & Germany & $A_{9}$ & Luxembourg & $A_{13}$ & Slovak Republic \\
$A_{2}$ & Czech Republic & $A_{6}$ & Hungary & $A_{10}$ & Netherlands & $A_{14}$ & Slovenia \\
$A_{3}$ & Finland & $A_{7}$ & Iceland & $A_{11}$ & Norway & $A_{15}$ & Sweden \\
$A_{4}$ & France & $A_{8}$ & Latvia & $A_{12}$ & Poland & $A_{16}$ & United Kingdom \\ \bottomrule
\end{tabular}
}
\end{table}

Based on the framework and assumptions presented in section~\ref{sec:ConceptualFramework} regarding the completeness of public data made available by websites of OECD Health Statistics, Eurostat, and World Health Organization, accessed on 21 January 2022, after collecting the most recent data available and merging them, as a results authors received an evaluation decision matrix containing the performance values of 25 selected criteria for the sixteen alternatives evaluated in this study. This decision matrix is available to a broader audience in a repository available on GitHub~\citep{sspahpgithub2022}. In addition, the full Python code containing the framework of the performed research along with Step-by-step instructions with examples based on the problem presented in this article, complete input data and supporting methods, and the SSP-AHP method in a separate class is made available on GitHub, which thus can be easily used also for other decision problems than the problem presented in this paper. Table~\ref{tab:Criteria} provided in section~\ref{sec:ConceptualFramework} presents the brief summary of all considered criteria structured in the hierarchical model containing main criteria $G_{1}$--$G_{5}$ and sub-criteria in the health system’s social-sustainability assessment with names for main dimensions, subdimensions, detailed lowest level criteria, and objective of criteria, which is represented by 1 for profit criteria to be maximized and on the other hand -1 denotes objective to minimize for cost criteria. In our framework, we propose three sub-criteria to assess the health system’s adaptability: investments in public health, investments in human resources, and new technologies uptake. 

\begin{figure}[H]
    \centering
    \includegraphics[width=\linewidth]{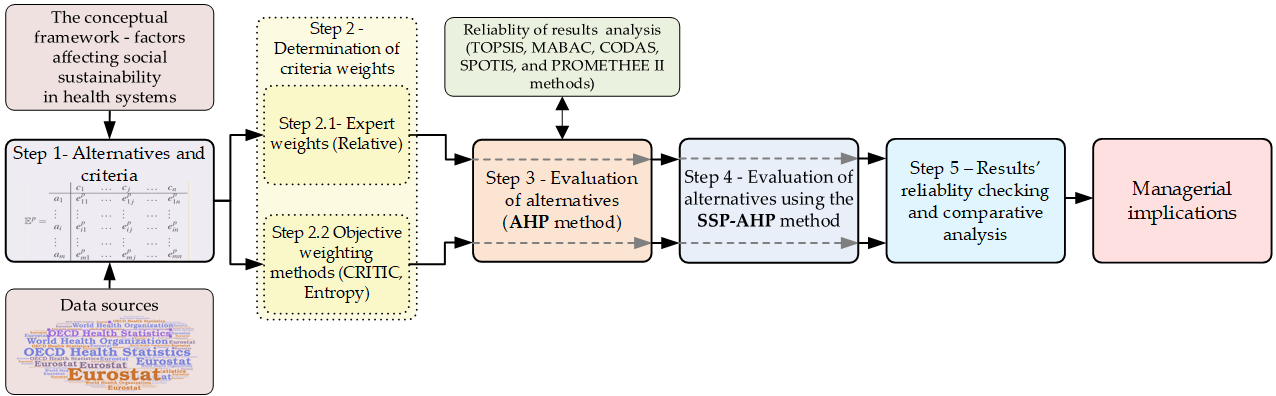}
    \caption{Flowchart presenting application of the SSP-AHP method in sustainable healthcare system assessment.}
    \label{fig:framework}
\end{figure}

Criteria weights were determined by experts using the AHP-based relative weighting method, which is described in detail in~\ref{sec:appWeights}. In addition, a comparative analysis was also performed using objective weighting methods for benchmarking purposes, namely Entropy~\citep{lotfi2010imprecise} and the Criteria Importance Through Inter criteria
Correlation (CRITIC)~\citep{tucs2019new} weighting methods. It should be noted primarily that these are methods for determining objective weights, which in our case results in the ability to determine the objective value of the indicator without the involvement of experts, and are often used by researchers in MCDA where there is a need to determine an objective MCDA ranking without the use of expert knowledge~\citep{abdel2020novel, du2020decision, li2020regional}. These techniques were chosen as reference methods, a justification of which can be found in~\citep{zhao2020comprehensive, rostamzadeh2018evaluation}. The CRITIC method has the advantage of considering two dimensions of information contained in the data when determining the weights. The first dimension is the contrast intensity between the criteria, measured using the standard deviation. The second dimension is the conflict between criteria measured using the linear correlation coefficient. Considering the two dimensions mentioned above improves the accuracy of multi-criteria decision analysis methods~\citep{li2022critic}. In the case of the Entropy weighting method, a significant difference between criterion values represents a high amount of information in the data. The lower the entropy value, the higher the criterion's weight. This method is based on Shannon entropy, which measures the uncertainty or variability contained in the data. The lower the entropy value, the lower the degree of chaos~\citep{li2022suitability, sidhu2022bibliometric}. Of course, expert values can also be used; hence our comparison with expert weights was determined using an AHP function called the eigenvector method~\citep{duleba2022introduction}. These objective weighting methods are detailed in~\ref{sec:appWeights}. In the first part of this research, a comparative analysis of the results provided by the classical AHP method with six benchmarking MCDA methods was conducted to confirm the reliable results of the AHP approach. Then, a comparative analysis with objective weighting methods was conducted to confirm the reliability of weights determined subjectively in the classical AHP approach. In the following research stage, an investigation using SSP-AHP increasing the compensation reduction under each main criteria was performed. The detailed research framework is presented in Figure~\ref{fig:framework}, and the individual steps of the framework are described in detail below.

\textbf{Step 1.} Selection of alternatives and criteria.

\noindent
The selection of the countries for social sustainability-oriented health systems analysis listed in Table~\ref{tab:countries} has been based on the healthcare financing model observed in Europa (Beverage model, Bismarck model, and hybrid one - the national health insurance model).

\textbf{Step 2.} Determination of criteria weights.

\noindent
In this research, to provide reliable results, the authors used weights determined subjectively with expert involvement and weights determined from performance values contained in the decision matrix using objective Entropy and CRITIC weighting methods in separate framework flows. For both approaches of determining weights, the rest of the framework steps are identical.

\textbf{Step 2.1} Determination of criteria weights with the subjective AHP-based relative weighting method involving experts.

\noindent
In the first part of the research, the weights for each criteria dimension, namely $G_{1}$-$G_{5}$, were determined based on the opinion of the domain expert using the scale proposed by Saaty. According to AHP methodology~\citep{rahman2021multi} small sample size is acceptable, as AHP involves expert opinion rather than surveys. In our case, the research sample consisted of sixteen experts, which is considered to be satisfactory and comparable to other studies in the field~\citep{aljaberi2017framework}. Experts in senior managerial positions at various healthcare entities and organizations (hospitals, insurance schemes, national health funds, ministry of health) from the surveyed countries have been invited to participate in an online questionnaire. The selected respondents had at least ten years of expert experience in the health sector. In addition, they all have been engaged in decision-making processes while setting and prosecuting health policies and strategies. Details of the experts and the methods used to recruit them for the study are provided in~\ref{app:experts}.

In the form of pairwise competitions, experts' judgments served as input information for AHP analysis. The structured AHP model expresses the social sustainability priority in respect of a set of relative importance weights for the five major categories of social-sustainable health systems. In line with AHP methodology~\citep{dehe2015development}, the geometric mean approach was preferred over the arithmetic mean to combine the individual pairwise comparison judgments of the 16 healthcare experts into the consensus pairwise comparison matrix. The criteria pairwise comparison matrix provided by an expert is displayed in Table~\ref{tab:practicalCriteriaComparison}.

\begin{table}[H]
\centering
\caption{Pairwise comparison matrix of criteria significance.}
\label{tab:practicalCriteriaComparison}
\resizebox{0.7\linewidth}{!}{
\begin{tabular}{cccccc} \toprule
 & Equity & Quality & Responsiveness & Financial coverage & Adaptability \\ \midrule
Equity & 1 & 1 & 5 & 3 & 9 \\
Quality & 1 & 1 & 3 & 5 & 7 \\
Responsiveness & 1/5 & 1/3 & 1 & 1 & 9 \\
Financial coverage & 1/3 & 1/5 & 1 & 1 & 7 \\
Adaptability & 1/9 & 1/7 & 1/9 & 1/7 & 1 \\ \bottomrule
\end{tabular}
}
\end{table}

The next stage was to check the consistency of the pairwise comparison matrix of criteria significance, which determines whether the pairwise relationship is consistent, using the $CR$ presented in Equation~(\ref{eq:CR}) coefficient. For this matrix, the $CR$ value is equal to 0.08. A value less than 0.1 means that the pairwise comparison result is acceptable, so this matrix containing pairwise comparison criteria is consistent. Then the vector of weights included in Table~\ref{tab:weightsMainSaaty} for criteria $G_{1}$-$G_{5}$ was determined using the eigenvector method proposed by Saaty.

\begin{table}[H]
\centering
\caption{Main criteria weights determined with the AHP-based relative weighting method.}
\label{tab:weightsMainSaaty}
\resizebox{0.5\linewidth}{!}{
\begin{tabular}{lrrrrr} \toprule
Main criterion & $G_{1}$ & $G_{2}$ & $G_{3}$ & $G_{4}$ & $G_{5}$ \\ \midrule
Weights & 0.3689 & 0.3546 & 0.1292 & 0.1191 & 0.0282 \\ \bottomrule
\end{tabular}
}
\end{table}
It can be observed that in the opinion of experts, the groups of criteria $G_{1}$ (Equity) and $G_{2}$ (Quality of care) turned out to be the most important, as they received the highest values of weights. The weights determined for the main criteria were then distributed equally to the sub-dimensions and particular criteria, as shown in Table~\ref{tab:weightsDistributedSaaty}.

\begin{table}[H]
\centering
\caption{Weights determined for dimensions, sub-dimensions, and particular criteria by distributing equally main criteria weights to sub-criteria.}
\label{tab:weightsDistributedSaaty}
\resizebox{0.6\linewidth}{!}{
\begin{tabular}{lrrr} \toprule
Criterion & Dimension weights  & Subdimension weights & Criterion weights \\ \midrule
$C_{1}$ & \multirow{7}{*}{0.3689} & \multirow{2}{*}{0.1230} & 0.0615 \\
$C_{2}$ &  &  & 0.0615 \\
$C_{3}$ &  & \multirow{2}{*}{0.1230} & 0.0615 \\
$C_{4}$ &  &  & 0.0615 \\
$C_{5}$ &  & \multirow{3}{*}{0.1230} & 0.0410 \\
$C_{6}$ &  &  & 0.0410 \\
$C_{7}$ &  &  & 0.0410 \\ \midrule
$C_{8}$ & \multirow{7}{*}{0.3546} & \multirow{4}{*}{0.1182} & 0.0296 \\
$C_{9}$ &  &  & 0.0296 \\
$C_{10}$ &  &  & 0.0296 \\
$C_{11}$ &  &  & 0.0296 \\
$C_{12}$ &  & 0.1182 & 0.1182 \\
$C_{13}$ &  & \multirow{2}{*}{0.1182} & 0.0591 \\
$C_{14}$ &  &  & 0.0591 \\ \midrule
$C_{15}$ & \multirow{3}{*}{0.1292} & 0.0646 & 0.0646 \\
$C_{16}$ &  & \multirow{2}{*}{0.0646} & 0.0323 \\
$C_{17}$ &  &  & 0.0323 \\ \midrule
$C_{18}$ & \multirow{3}{*}{0.1191} & 0.0595 & 0.0595 \\
$C_{19}$ &  & \multirow{2}{*}{0.0595} & 0.0298 \\
$C_{20}$ &  &  & 0.0298 \\ \midrule
$C_{21}$ & \multirow{5}{*}{0.0282} & \multirow{2}{*}{0.0094} & 0.0047 \\
$C_{22}$ &  &  & 0.0047 \\
$C_{23}$ &  & \multirow{2}{*}{0.0094} & 0.0047 \\
$C_{24}$ &  &  & 0.0047 \\
$C_{25}$ &  & 0.0094 & 0.0094 \\ \midrule
\end{tabular}
}
\end{table}

\textbf{Step 2.2} Determination of criteria weights using two objective weighting methods: Entropy and CRITIC weighting methods.

\noindent
To confirm the reliability of results obtained applying the relative weighting method with expert participation, weights were also determined with two objective weighting methods for comparative analysis, namely Entropy and CRITIC weighting. These objective weighting methods are based on performance values in the dataset, have proven their usefulness in many studies involving MCDA assessment, and are well-established alternatives to expert weights.

The obtained values of the weights by both objective weighting techniques show that the CRITIC method more closely reflects the values of the weights determined subjectively with the experts since it also identified criteria $G_{1}$ and $G_{2}$ as the most relevant. Thus, it can be assumed that the rankings provided by the SSP-AHP method using expert weights will be more consistent with the results obtained using the weights determined by the CRITIC method than Entropy. Criteria weights determined by all the weighting methods used in this paper are included in Table~\ref{tab:resultsWeights} in~\ref{app:resultsWeights}.

\textbf{Step 3.} MCDA evaluation of alternatives using the SSP-AHP method without criteria compensation reduction.

\noindent
In the following research stage, the ranking of the SSP-AHP method is generated for the subjective weights of the criteria without reduction of criteria compensation. The SSP-AHP method applied without criteria compensation reduction is equivalent to classical AHP. To confirm results' reliability, obtained ranking is compared with rankings provided by benchmarking methods using correlation coefficients. The set of benchmarking methods includes well-established and widely used MCDA methods, namely TOPSIS~\citep{chmielarz2022assessment}, MABAC~\citep{pamuvcar2015selection}, CODAS~\citep{badi2018site}, SPOTIS~\citep{dezert2020spotis}, and PROMETHEE II~\citep{salabun2020mcda, sotiropoulou2021onshore}. The basics, algorithmic assumptions of these methods, and details of their steps can be found in the given references. The motivation for selecting the indicated set of reference methods was their several advantages. The PROMETHEE II method belongs to the European methods and is an outranking method with low criteria compensation~\citep{sotiropoulou2021onshore}. The other methods represent the American school. The MABAC method is characterized by high stability for changes in the model, for example, within weights~\citep{pamuvcar2015selection}. The TOPSIS method is a popular and established method that measures the Euclidean distance of alternatives from reference points, which are the ideal and anti-ideal solutions, to determine the best alternative~\citep{chmielarz2022assessment}. The CODAS method, like TOPSIS, evaluates alternatives based on the distance from the reference points, but it uses the Euclidean and taxicab distances from the anti-ideal solution~\citep{badi2018site}. The SPOTIS method also uses the concept of reference objects. However, instead of reference objects, it uses performance value boundaries as the ideal and anti-ideal solution, which prevents ranking reversal in the case of adding or removing an alternative from the evaluated set~\citep{dezert2020spotis}.

Then, also to confirm the reliability of results, the convergence of the SSP-AHP rankings obtained for subjective weights of criteria is compared with SSP-AHP rankings obtained for objective weights of criteria. The correlation of SSP-AHP ranking with the rankings of reference methods for objective weights is also examined.

\textbf{Step 4.} MCDA evaluation of alternatives using the SSP-AHP method with criteria compensation reduction.

\noindent
The following research step is to apply the SSP-AHP method with a stepwise increasing value of sustainability coefficient, gradually reducing the criteria compensation. The procedure is performed for the expert criteria weights and to confirm the reliability of results for the objective weights. The convergence of the results is measured using two ranking correlation coefficients: Weighted Spearman's Rank Correlation Coefficient ($r_w$)~\citep{salabun2020mcda}, and Pearson coefficient~\citep{deng2021combining} detailed in~\ref{sec:appCorrelations}.

\textbf{Step 5.} Comparative analysis of results with a set of benchmark MCDA and weighting methods.

\noindent
For this purpose, the rankings were compared using two measures of correlation, namely the $r_w$ and Pearson coefficient.

\textbf{Step 6. } Formulating managerial implications. 

\noindent
This step includes determination based on the received rankings, which countries are the most sustainable and take the leading positions in terms of the sustainable healthcare system.
Finally, implications for the sustainability of the healthcare system of analyzed countries are generated with a particular focus on the effect of reducing the degree of criteria compensation on the final rankings.

\section{Results}
\label{sec:results}
This section presents and discusses the results of multi-criteria evaluating the alternatives in terms of a sustainable healthcare system. For this purpose, with the help of SSP-AHP without criteria compensation reduction using the criteria weights determined by the AHP-based relative weighting method, a multi-criteria evaluation of alternatives was carried out, resulting in utility function values. For the SSP-AHP method, the highest utility function value score indicates the best-scored alternative. Rankings were generated concerning the sustainable healthcare system by sorting the utility function values in descending order. Then the received rankings to confirm their reliability were compared with the rankings obtained using a set of five benchmarking MCDA methods. The SSP-AHP results were also compared with the results gained for two different objective weighting methods: Entropy and CRITIC weighting methods, to confirm the reliability of weights determined with the AHP-based relative weighting method involving experts. The next stage of the research involved a stepwise reduction of criteria compensation in the SSP-AHP method by gradually increasing the value of $s$ sustainability coefficient. This stage aimed to identify the most sustainable countries in terms of the healthcare system, which remain stable despite reducing criteria compensation. This procedure was performed for weights determined with AHP-based relative weighting method involving experts, as well as for weights calculated with objective weighting methods to confirm the reliability of AHP-based relative weights.

\subsection{Results of the SSP-AHP evaluation with using weights determined with AHP-based relative weighting method}

\begin{figure}[H]
    \centering
    \includegraphics[width=0.7\linewidth]{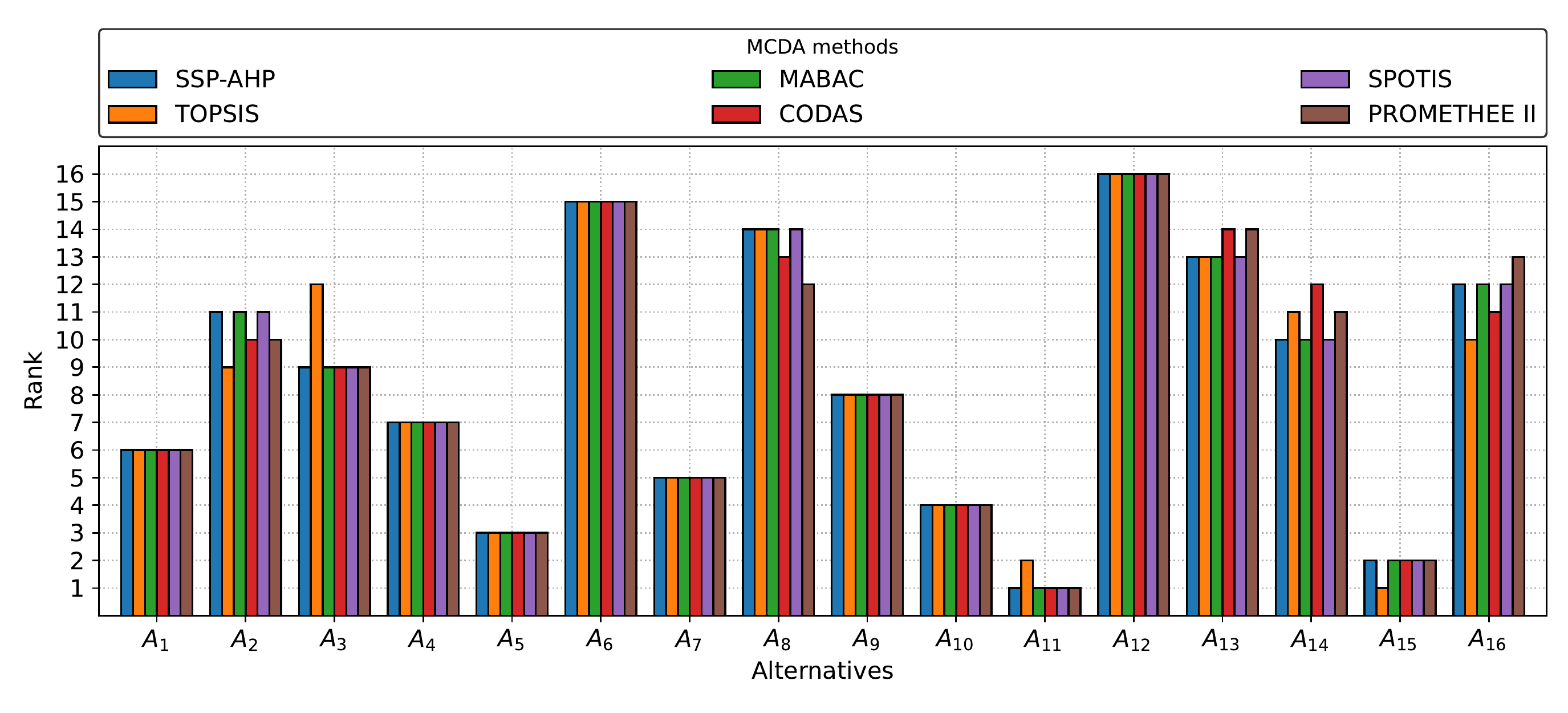}
    \caption{Comparison of rankings provided by the SSP-AHP method without reduction of criteria compensation and reference MCDA methods for criteria weights determined with AHP-based relative weighting method.}
    \label{fig:resultsBarChartAHP}
\end{figure}

\begin{figure}[H]
\centering
\includegraphics[width=0.45\linewidth]{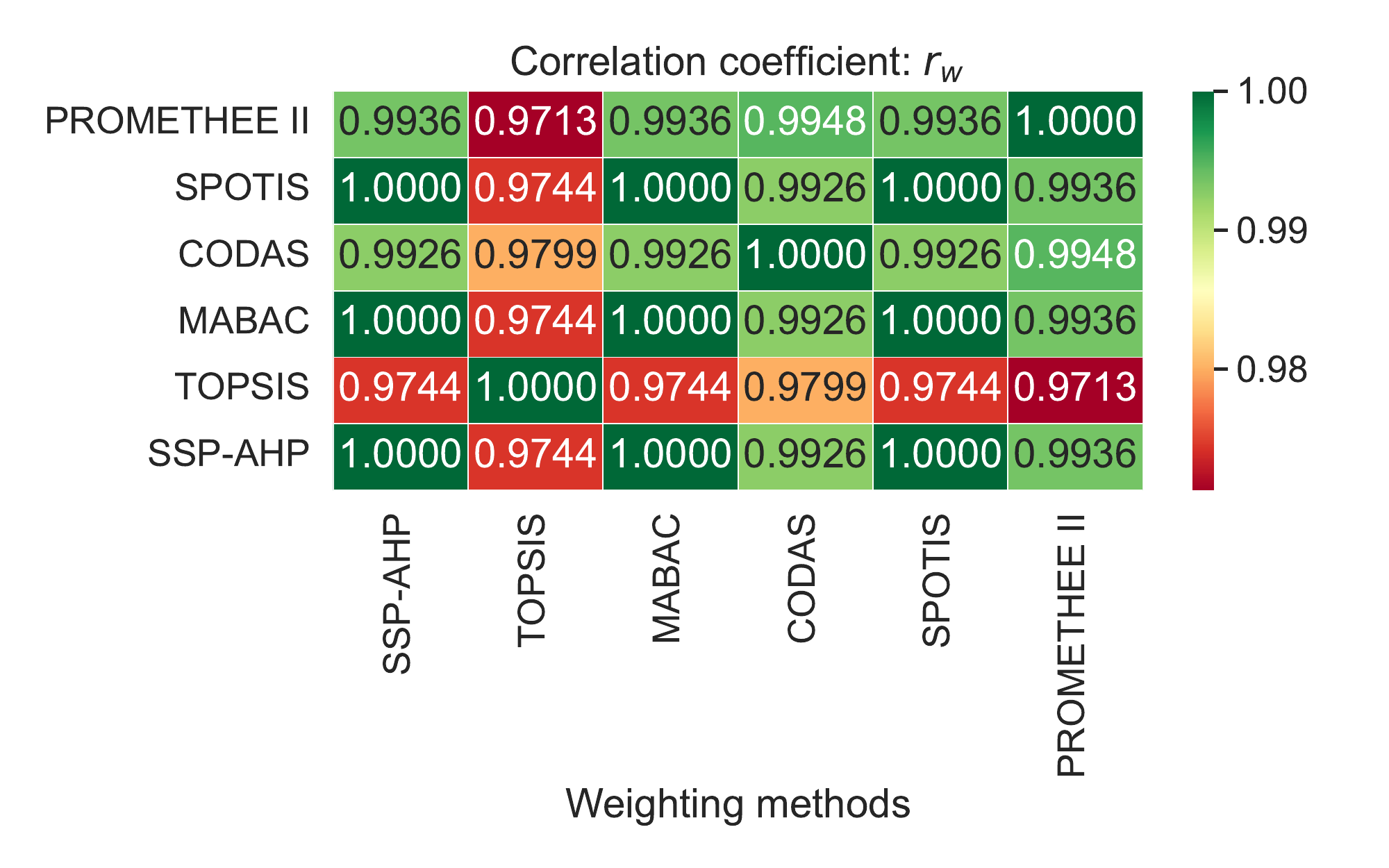}
\includegraphics[width=0.45\linewidth]{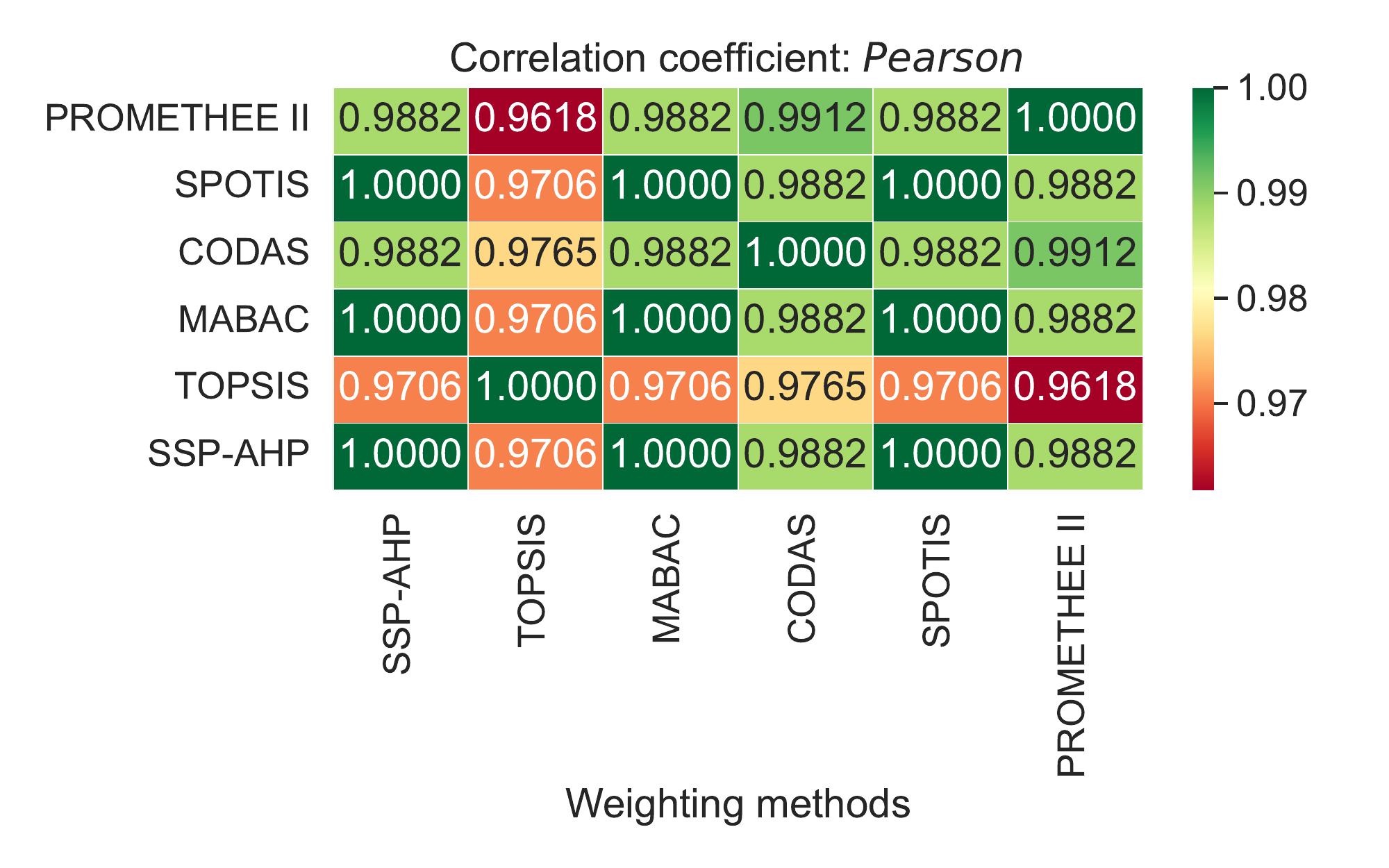}
\caption{Correlations of rankings provided by the SSP-AHP method without criteria compensation reduction and other MCDA methods for criteria weights determined with AHP-based relative weighting method.}
\label{fig:correlationsAHPSaaty}
\end{figure}

The essence of the SSP-AHP method, as with most MCDA methods, is to generate a quantitative ranking of the evaluated alternatives. First, a ranking was generated for the SSP-AHP method without criterion compensation reduction using criterion weights determined by the AHP-based relative weighting method involving experts, which was compared with rankings obtained by applying five benchmark MCDA methods to confirm the results reliability and relevance of the AHP method to the problem under investigation. The results of the SSP-AHP method without reduction of criteria compensation and five other multi-criteria methods selected as reference methods, namely TOPSIS MABAC, CODAS, SPOTIS, and PROMETHEE II, are presented in Table~\ref{tab:resultsSaaty}. The provided table contains the utility function values and rankings for each alternative. Received rankings were also visualized in the column chart shown in Figure~\ref{fig:resultsBarChartAHP}.
\begin{table}[H]
\caption{Utility function values and rankings provided by the SSP-AHP method without criteria compensation reduction and reference methods using criteria weights determined with AHP-based relative weighting method.}
\label{tab:resultsSaaty}
\resizebox{\linewidth}{!}{
\begin{tabular}{llrrrrrrrrrrrr} \toprule
$A_{i}$ & Country & SSP-AHP & TOPSIS & MABAC & CODAS & SPOTIS & PROMETHEE II & SSP-AHP & TOPSIS & MABAC & CODAS & SPOTIS & PROMETHEE II \\ \midrule
$A_{1}$ & Belgium & 0.6243 & 0.5530 & 0.1199 & 1.2297 & 0.3757 & 0.1164 & 6 & 6 & 6 & 6 & 6 & 6 \\
$A_{2}$ & Czech Republic & 0.4920 & 0.4852 & -0.0124 & -0.9183 & 0.5080 & -0.1216 & 11 & 9 & 11 & 10 & 11 & 10 \\
$A_{3}$ & Finland & 0.5072 & 0.4579 & 0.0028 & -0.5450 & 0.4928 & -0.0645 & 9 & 12 & 9 & 9 & 9 & 9 \\
$A_{4}$ & France & 0.5776 & 0.5528 & 0.0732 & 0.4678 & 0.4224 & 0.0962 & 7 & 7 & 7 & 7 & 7 & 7 \\
$A_{5}$ & Germany & 0.6775 & 0.6385 & 0.1731 & 2.7486 & 0.3225 & 0.3337 & 3 & 3 & 3 & 3 & 3 & 3 \\
$A_{6}$ & Hungary & 0.3272 & 0.4003 & -0.1772 & -3.5673 & 0.6728 & -0.3599 & 15 & 15 & 15 & 15 & 15 & 15 \\
$A_{7}$ & Iceland & 0.6338 & 0.5671 & 0.1294 & 1.7713 & 0.3662 & 0.1934 & 5 & 5 & 5 & 5 & 5 & 5 \\
$A_{8}$ & Latvia & 0.3589 & 0.4444 & -0.1455 & -1.6453 & 0.6411 & -0.2231 & 14 & 14 & 14 & 13 & 14 & 12 \\
$A_{9}$ & Luxembourg & 0.5360 & 0.5403 & 0.0316 & 0.1882 & 0.4640 & 0.0332 & 8 & 8 & 8 & 8 & 8 & 8 \\
$A_{10}$ & Netherlands & 0.6391 & 0.6257 & 0.1347 & 2.0197 & 0.3609 & 0.2486 & 4 & 4 & 4 & 4 & 4 & 4 \\
$A_{11}$ & Norway & 0.7311 & 0.6408 & 0.2267 & 3.6190 & 0.2689 & 0.4893 & 1 & 2 & 1 & 1 & 1 & 1 \\
$A_{12}$ & Poland & 0.3132 & 0.3454 & -0.1912 & -4.3183 & 0.6868 & -0.4865 & 16 & 16 & 16 & 16 & 16 & 16 \\
$A_{13}$ & Slovak Republic & 0.4001 & 0.4456 & -0.1043 & -2.1426 & 0.5999 & -0.2830 & 13 & 13 & 13 & 14 & 13 & 14 \\
$A_{14}$ & Slovenia & 0.5020 & 0.4779 & -0.0024 & -1.1802 & 0.4980 & -0.1422 & 10 & 11 & 10 & 12 & 10 & 11 \\
$A_{15}$ & Sweden & 0.6863 & 0.6625 & 0.1819 & 3.2536 & 0.3137 & 0.4183 & 2 & 1 & 2 & 2 & 2 & 2 \\
$A_{16}$ & United Kingdom & 0.4813 & 0.4806 & -0.0231 & -0.9808 & 0.5187 & -0.2483 & 12 & 10 & 12 & 11 & 12 & 13 \\ \bottomrule
\end{tabular}
}
\end{table}
In order to confirm the reliability of the SSP-AHP result, the authors checked the convergence of the SSP-AHP ranking with the other rankings provided by the reference methods. Two coefficients, namely Pearson's correlation coefficient and Weighted Spearman's Rank Correlation Coefficient $r_w$, were used to measure rankings' convergence. The values of the calculated correlations between the rankings are visualized in Figure~\ref{fig:correlationsAHPSaaty}. The obtained results show that the rankings identical to those provided by the SSP-AHP method gave the MABAC and SPOTIS methods. Besides, strong correlation values were observed when comparing the rankings provided by the SSP-AHP method with those of PROMETHEE II and CODAS. In the case of TOPSIS, the correlation values were the lowest of all obtained reference rankings. However, their values are still high. The ranking leader of the SSP-AHP and the other reference methods except TOPSIS is Norway ($A_{11}$). For the TOPSIS method ranking, the first place went to Sweden. For the second place TOPSIS ranking, it is exactly in the opposite order, i.e., for the AHP and the other benchmark methods except for TOPSIS, Sweden ($A_{15}$) is second, and for the TOPSIS ranking, Norway is second. Then from the third place until eighth place, the rankings of SSP-AHP and all benchmark methods are identical. It is easy to notice that the countries that received the best scores are Norway, Sweden, Germany ($A_{5}$), the Netherlands ($A_{10}$), and Iceland ($A_{7}$).
%

\begin{figure}[H]
    \centering
    \includegraphics[width=0.5\linewidth]{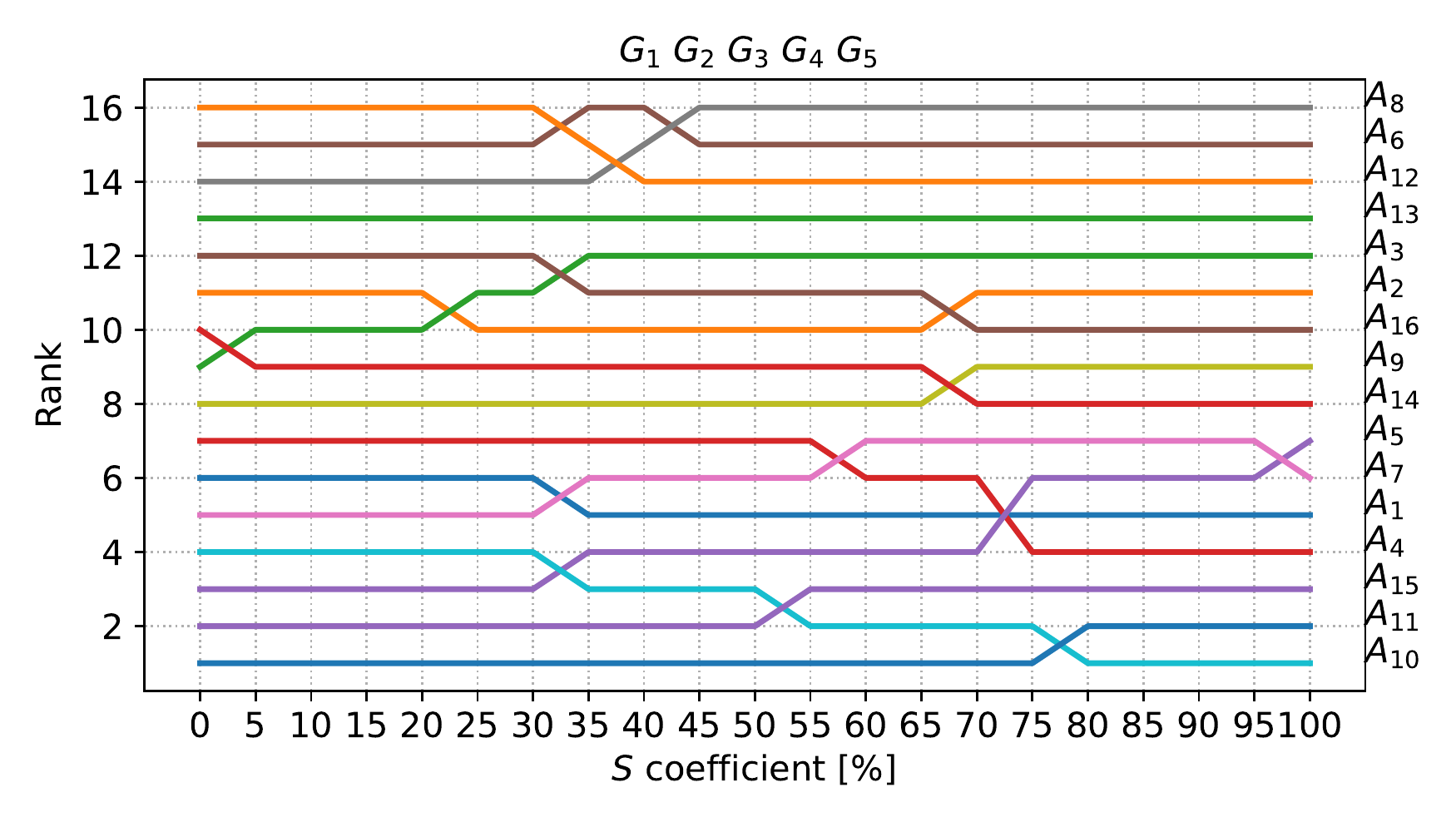}
    \caption{Changes in rankings with modification of $s$ coefficient in all criteria for criteria weights determined with the SSP-AHP method.}
    \label{fig:sustCoeffSaaty}
\end{figure}

On the other hand, the worst-ranked countries in all the rankings are Hungary in the second-to-last place and Poland ($A_{12}$) in the last place. In the case of the TOPSIS method, the number of different places and the range of differences are higher than in the case of SSP-AHP comparison with other reference methods, which is reflected in the form of lower values of correlation coefficients. In conclusion, the determined correlations are sufficiently high to confirm the high convergence of all rankings, which proves the reliability of SSP-AHP in the investigated problem.
%
%
In the next part of the research to model the strength of the criteria compensation reduction with the SSP-AHP method, an analysis of increasing the value of sustainability coefficient $s$ in increments of 5\% starting from a value of 0\% up to 100\% was performed for criteria belonging to all possible combinations of the main dimensions, thus reducing the criteria compensation progressively. An increase of sustainability factor is equal to the stepwise reduction of criteria compensation starting from full criteria compensation ($s$ equal to 0\%) and ending with full criteria compensation reduction ($s$ equal to 100\%). Table~\ref{tab:sustSSPAHPweightSaaty} and Figure~\ref{fig:sustCoeffSaaty} demonstrates the changes in rankings while increasing the value of $s$ for all criteria dimensions.

\begin{table}[H]
\centering
\caption{SSP-AHP rankings for criteria compensation reduction with $s$ coefficient for all combinations of criteria dimensions $G_{1}$--$G_{5}$.}
\label{tab:sustSSPAHPweightSaaty}
\resizebox{0.65\linewidth}{!}{
\begin{tabular}{lrrrrrrrrrrrrrrrr} \toprule
G & $A_{1}$ & $A_{2}$ & $A_{3}$ & $A_{4}$ & $A_{5}$ & $A_{6}$ & $A_{7}$ & $A_{8}$ & $A_{9}$ & $A_{10}$ & $A_{11}$ & $A_{12}$ & $A_{13}$ & $A_{14}$ & $A_{15}$ & $A_{16}$ \\ \midrule
 & 6 & 11 & 9 & 7 & 3 & 15 & 5 & 14 & 8 & 4 & 1 & 16 & 13 & 10 & 2 & 12 \\
$G_{5}$ & 6 & 11 & 9 & 7 & 3 & 15 & 5 & 14 & 8 & 4 & 1 & 16 & 13 & 10 & 2 & 12 \\
$G_{4}$ & 5 & 11 & 9 & 7 & 6 & 15 & 3 & 14 & 8 & 4 & 1 & 16 & 13 & 10 & 2 & 12 \\
$G_{4}$ $G_{5}$ & 5 & 11 & 9 & 7 & 6 & 15 & 3 & 14 & 8 & 4 & 1 & 16 & 13 & 10 & 2 & 12 \\
$G_{3}$ & 7 & 11 & 10 & 6 & 2 & 14 & 5 & 15 & 8 & 4 & 1 & 16 & 13 & 9 & 3 & 12 \\
$G_{3}$ $G_{5}$ & 7 & 11 & 10 & 6 & 2 & 14 & 5 & 15 & 8 & 4 & 1 & 16 & 13 & 9 & 3 & 12 \\
$G_{3}$ $G_{4}$ & 6 & 11 & 10 & 7 & 3 & 14 & 4 & 15 & 8 & 5 & 1 & 16 & 13 & 9 & 2 & 12 \\
$G_{3}$ $G_{4}$ $G_{5}$ & 6 & 11 & 10 & 7 & 3 & 14 & 4 & 15 & 8 & 5 & 1 & 16 & 13 & 9 & 2 & 12 \\
$G_{2}$ & 5 & 11 & 12 & 7 & 2 & 16 & 6 & 15 & 8 & 3 & 1 & 14 & 13 & 9 & 4 & 10 \\
$G_{2}$ $G_{5}$ & 5 & 11 & 12 & 7 & 2 & 16 & 6 & 15 & 8 & 3 & 1 & 14 & 13 & 9 & 4 & 10 \\
$G_{2}$ $G_{4}$ & 4 & 11 & 12 & 7 & 2 & 16 & 6 & 15 & 8 & 3 & 1 & 14 & 13 & 10 & 5 & 9 \\
$G_{2}$ $G_{4}$ $G_{5}$ & 4 & 11 & 12 & 7 & 2 & 16 & 6 & 15 & 8 & 3 & 1 & 14 & 13 & 10 & 5 & 9 \\
$G_{2}$ $G_{3}$ & 7 & 11 & 12 & 5 & 2 & 15 & 6 & 16 & 9 & 3 & 1 & 14 & 13 & 8 & 4 & 10 \\
$G_{2}$ $G_{3}$ $G_{5}$ & 7 & 11 & 12 & 5 & 2 & 15 & 6 & 16 & 9 & 3 & 1 & 14 & 13 & 8 & 4 & 10 \\
$G_{2}$ $G_{3}$ $G_{4}$ & 7 & 11 & 12 & 5 & 2 & 15 & 6 & 16 & 9 & 3 & 1 & 14 & 13 & 8 & 4 & 10 \\
$G_{2}$ $G_{3}$ $G_{4}$ $G_{5}$ & 7 & 11 & 12 & 5 & 2 & 15 & 6 & 16 & 9 & 3 & 1 & 14 & 13 & 8 & 4 & 10 \\
$G_{1}$ & 4 & 10 & 11 & 7 & 6 & 15 & 5 & 14 & 8 & 2 & 3 & 16 & 13 & 9 & 1 & 12 \\
$G_{1}$ $G_{5}$ & 4 & 10 & 11 & 7 & 6 & 15 & 5 & 14 & 8 & 2 & 3 & 16 & 13 & 9 & 1 & 12 \\
$G_{1}$ $G_{4}$ & 4 & 10 & 11 & 6 & 7 & 15 & 5 & 14 & 8 & 2 & 3 & 16 & 13 & 9 & 1 & 12 \\
$G_{1}$ $G_{4}$ $G_{5}$ & 4 & 10 & 11 & 6 & 7 & 15 & 5 & 14 & 8 & 2 & 3 & 16 & 13 & 9 & 1 & 12 \\
$G_{1}$ $G_{3}$ & 7 & 10 & 11 & 6 & 4 & 14 & 5 & 15 & 8 & 2 & 3 & 16 & 13 & 9 & 1 & 12 \\
$G_{1}$ $G_{3}$ $G_{5}$ & 7 & 10 & 11 & 6 & 4 & 14 & 5 & 15 & 8 & 2 & 3 & 16 & 13 & 9 & 1 & 12 \\
$G_{1}$ $G_{3}$ $G_{4}$ & 6 & 10 & 11 & 5 & 7 & 14 & 4 & 15 & 8 & 2 & 3 & 16 & 13 & 9 & 1 & 12 \\
$G_{1}$ $G_{3}$ $G_{4}$ $G_{5}$ & 6 & 10 & 11 & 5 & 7 & 14 & 4 & 15 & 8 & 2 & 3 & 16 & 13 & 9 & 1 & 12 \\
$G_{1}$ $G_{2}$ & 4 & 10 & 12 & 6 & 5 & 15 & 7 & 16 & 8 & 1 & 2 & 14 & 13 & 9 & 3 & 11 \\
$G_{1}$ $G_{2}$ $G_{5}$ & 4 & 10 & 12 & 6 & 5 & 15 & 7 & 16 & 8 & 1 & 2 & 14 & 13 & 9 & 3 & 11 \\
$G_{1}$ $G_{2}$ $G_{4}$ & 3 & 11 & 12 & 5 & 7 & 16 & 6 & 15 & 8 & 1 & 2 & 14 & 13 & 9 & 4 & 10 \\
$G_{1}$ $G_{2}$ $G_{4}$ $G_{5}$ & 3 & 11 & 12 & 5 & 7 & 15 & 6 & 16 & 8 & 1 & 2 & 14 & 13 & 9 & 4 & 10 \\
$G_{1}$ $G_{2}$ $G_{3}$ & 6 & 10 & 12 & 5 & 4 & 15 & 7 & 16 & 9 & 1 & 2 & 14 & 13 & 8 & 3 & 11 \\
$G_{1}$ $G_{2}$ $G_{3}$ $G_{5}$ & 6 & 10 & 12 & 5 & 4 & 15 & 7 & 16 & 9 & 1 & 2 & 14 & 13 & 8 & 3 & 11 \\
$G_{1}$ $G_{2}$ $G_{3}$ $G_{4}$ & 5 & 11 & 12 & 4 & 6 & 15 & 7 & 16 & 9 & 1 & 2 & 14 & 13 & 8 & 3 & 10 \\
$G_{1}$ $G_{2}$ $G_{3}$ $G_{4}$   $G_{5}$ & 5 & 11 & 12 & 4 & 7 & 15 & 6 & 16 & 9 & 1 & 2 & 14 & 13 & 8 & 3 & 10 \\ \bottomrule
\end{tabular}
}
\end{table}
The most significant decrease with increasing $s$ coefficient was observed for Germany ($A_{5}$). Germany ranked third for $s$ between 0 and 30\%. Then, when $s$ values were 35-70\%, it took fourth place, for $s$ in the 75-95\% range it dropped to sixth place, and for the full compensation reduction implied by $s$ equal 100\%, this country decreased to seventh place. A fall of 3 positions from 9th to 12th place with increasing $s$ coefficient was noticed for Finland ($A_{3}$). When considering the decreases, it is worth noting Iceland ($A_{7}$) (down two positions) from the 5th to seventh place, but with the last step of increasing $s$ from 95\% to 100\%, it finally moved up to the sixth place. Latvia ($A_{8}$) also dropped two places from 14th to the last 16th. Besides drops, promotions towards better places in the ranking were also observed as the value of $s$ was increased. For example, France ($A_{4}$) moved from 7th to fourth place. Netherlands ($A_{10}$) moved up significantly from fourth to first place. Poland ($A_{12}$) moved up from 16th to 14th, Slovenia ($A_{14}$) from 10th to 8th, and the United Kingdom ($A_{16}$) from 12th to 10th.
%

\begin{figure}[H]
    \centering
    \includegraphics[width=0.5\linewidth]{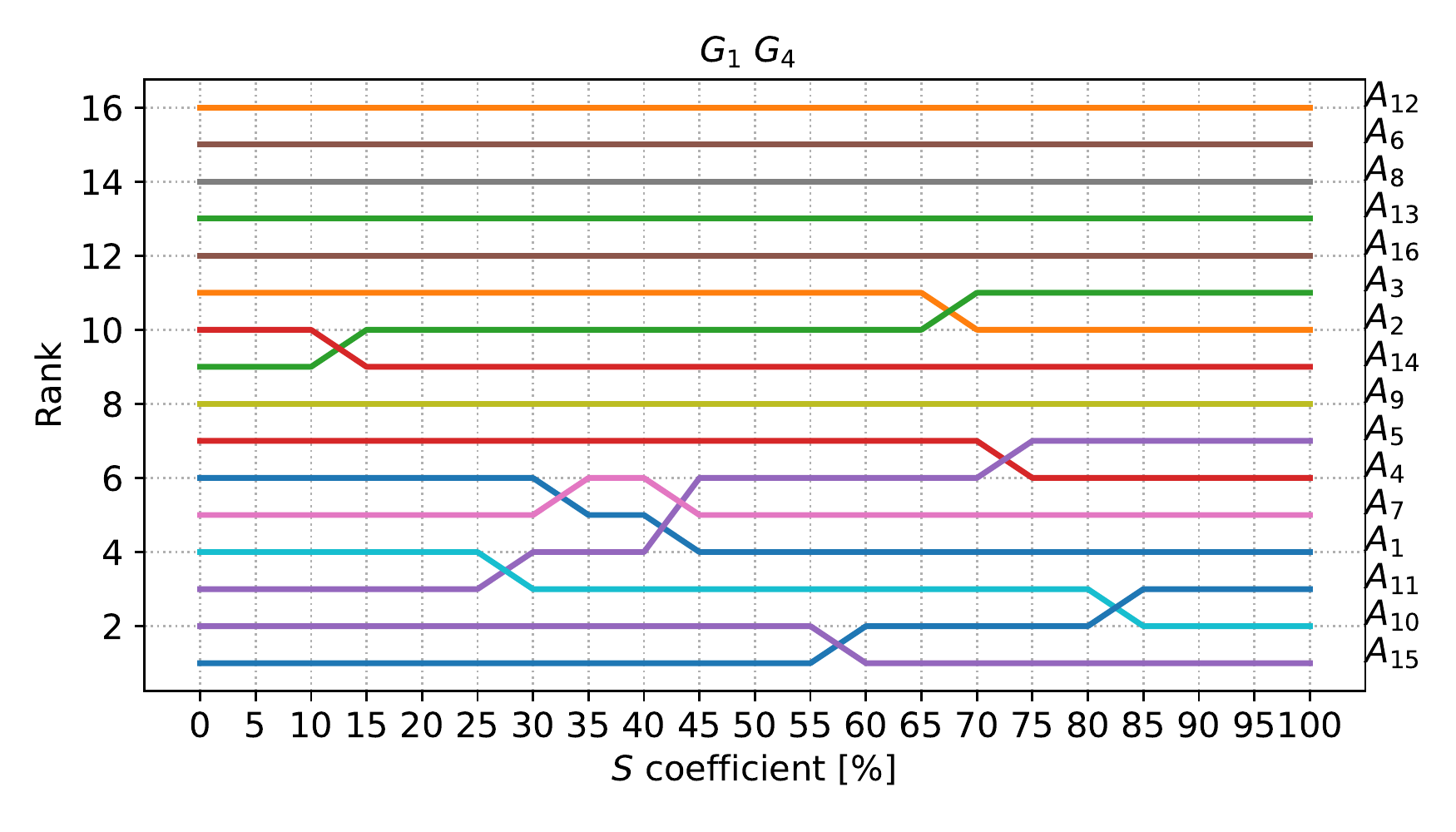}
    \caption{Demonstration of the decrease in Germany ($A_{5}$) in the SSP-AHP ranking as the $s$ coefficient increases for the criteria belonging to the main criteria $G_{1}$ and $G_{4}$.}
    \label{fig:sustCoeffSaaty18}
\end{figure}

\begin{figure}[H]
    \centering
    \includegraphics[width=0.5\linewidth]{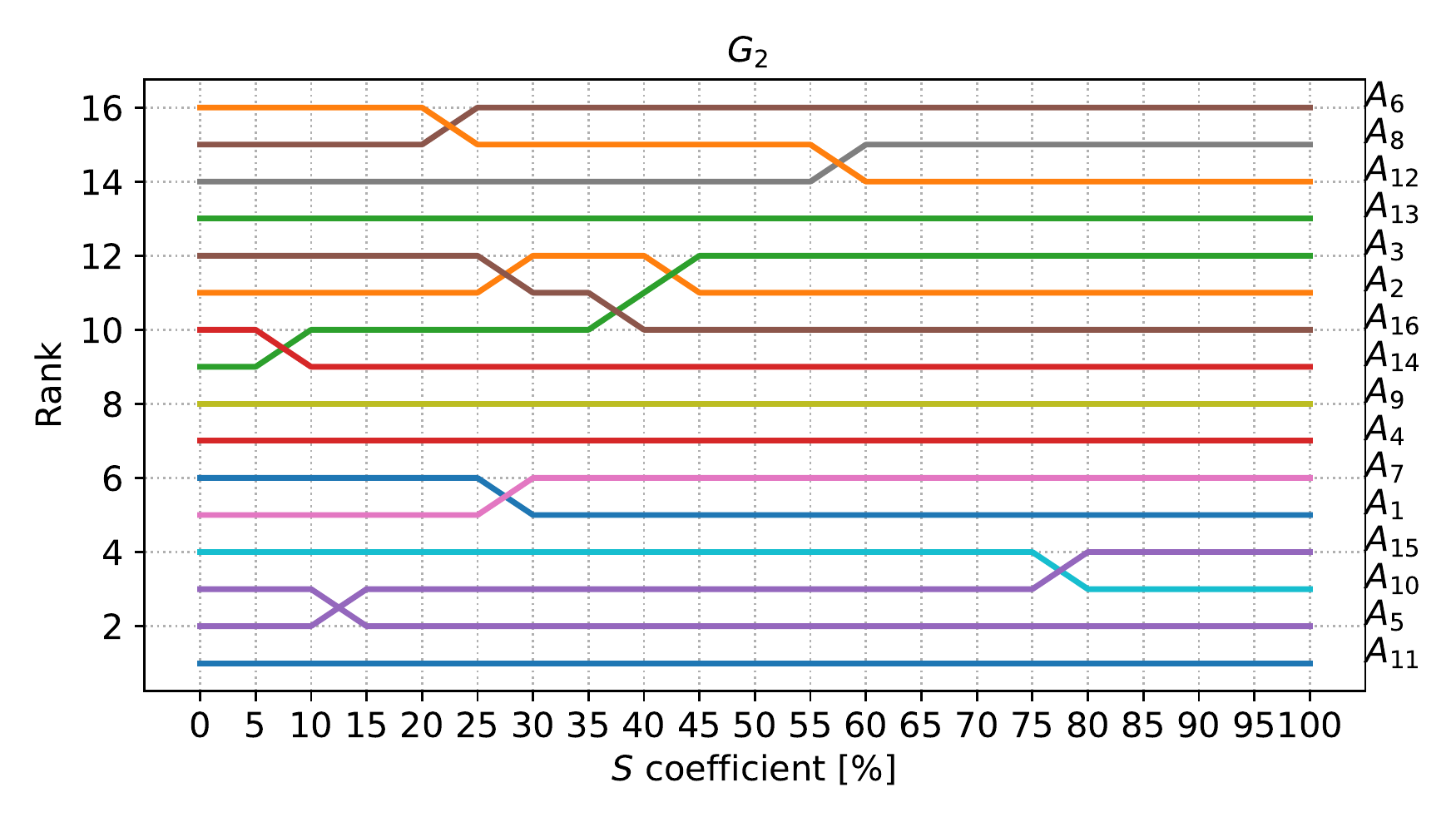}
    \caption{Demonstration of Germany's ($A_{5}$) advancement in the SSP-AHP ranking as the $s$ coefficient increases for the criteria belonging to the main criterion $G_{2}$.}
    \label{fig:sustCoeffSaaty8}
\end{figure}

An analogous analysis was then conducted by increasing the sustainability factor among all 32 possible combinations of 5 main dimensions (groups) of criteria. The simulations were performed to identify countries that demonstrate the greatest variability, that is, reducing compensation causes them to move up or down, and countries with high stability, including leaders and countries in the last positions of the rankings. Germany has shown the most significant susceptibility to a gradual reduction in criteria compensation. In the case of an unchanged $s$ coefficient, this country ranks third. However, during the sustainability analysis, the reduction in the compensation of criteria belonging to dimensions $G_{1}$ (Equity) in conjunction with $G_{4}$ (Financial coverage) was enough to register a drop of 5 positions for this country to seventh place. It is denoted in the graph displayed in Figure~\ref{fig:sustCoeffSaaty18}. 
%
However, when reducing the compensation of criteria belonging to dimensions $G_{2}$ or $G_{3}$, Germany moved up from third to second place, as shown in the graph in Figure~\ref{fig:sustCoeffSaaty8}. 
%
In the case of criterion dimension $G_{1}$ for Germany, criteria $C_{2}$ (Hip replacement), $C_{5}$ (acute care beds), $C_{7}$ (number of MRI machines) have the best, outstanding values of all analyzed countries, so it implies that they can compensate for lower values of other criteria. Thus, their ranking decreases when this compensation is reduced by increasing the $s$ coefficient. On the other hand, promotion is often observed when the $s$ is increased in the $G_{2}$ (Quality of care) and $G_{3}$ (Responsiveness) criteria groups.

Sweden ($A_{15}$) is another country for which variability in ranking position was noted during the reduction in criteria compensation. Sweden dropped from second to fifth place after increasing the coefficient $s$ for dimensions $G_{2}$ in conjunction with $G_{4}$. For $G_{2}$, Sweden has the best value for $C_{12}$ (\% of hospitalized patients with healthcare-associated infection) and $C_{13}$ (life expectancy). For $G_{4}$, the compensatory criterion for Sweden is $C_{18}$ (social health insurance in \% of the total population). In contrast, Sweden is promoted to the leader position as the coefficient $s$ increases in, for example, $G_{1}$ with no reduction in compensation in $G_{2}$. The variability in ranking due to decreasing compensation for criteria was also observed for Belgium ($A_{1}$), which rises from sixth to third place after decreasing compensation for dimensions $G_{1}$ in conjunction with $G_{2}$ and $G_{4}$. However, Belgium drops to seventh place when criteria compensation in dimension $G_{3}$ is reduced. Nevertheless, during the analysis, promotions were observed more often than decreases and were more significant, allowing Belgium to be considered a sustainable country.

Another country that rises or drops in the ranking depending on which criteria compensation is reduced for is Iceland ($A_{7}$). Iceland rises from fifth to third place when compensation is reduced for criteria in $G_{4}$ but decreases to seventh place when compensation is reduced for dimensions $G_{1}$ in conjunction with $G_{2}$. Netherlands ($A_{10}$) jumps from fourth to first place when compensation for criteria groups $G_{1}$ in conjunction with $G_{2}$ is reduced and falls to fifth place when compensation in the criteria group $G_{3}$ in conjunction with $G_{4}$ is reduced. Finland drops from 9th to 12th place when criteria compensation is reduced, especially $G_{2}$. For Finland, no promotions were observed with reduced criteria compensation during the analysis.

In contrast to Finland, France ($A_{4}$) is the country for which only promotions were observed when analyzing with increasing reductions in criteria compensation. The most significant advancement was observed for France when decreasing compensation for criteria belonging to conjoined dimensions $G_{1}$, $G_{2}$, $G_{3}$, and $G_{4}$ and for all dimensions. France then moved up from seventh to fourth place. Norway ($A_{11}$) takes the leading position in the ranking of the SSP-AHP method without criteria compensation reduction and the rankings of the four reference methods. When reducing compensation, decreases to the second position were observed when reducing compensation in criteria groups $G_{1}$ in conjunction with $G_{2}$ and to the third place when reducing compensation in criteria group $G_{1}$. It is worth noting that Norway remains in the first place when we do not limit the compensation in $G_{1}$. 

Norway has the best value for $C_{3}$ (practicing physicians) and $C_{4}$ (practicing nurses), and these criteria compensate other criteria that do not have as good values as the other countries. The leading position is maintained when there are no changes within the group of $G_{1}$ criteria. Besides, Norway has the highest compensating values for $C_{16}$, $C_{17}$ belonging to $G_{3}$, and $C_{18}$, $C_{19}$, $C_{20}$ involved in $G_{4}$. The country that has kept its position stable during the criteria compensation reduction analysis is the Slovak Republic ($A_{13}$). As can be observed, this country occupies a far thirteenth place in the ranking. The stability of this country in the ranking is because, for any criterion, it does not achieve an outstanding performance value, which could compensate for the worse values achieved for other criteria. Thus, the sustainability of this country finally results in stability and robustness of its rank evident during the analysis.

The worst-ranked countries are Hungary ($A_{6}$), Latvia ($A_{8}$), and Poland ($A_{12}$). Hungary drops from 15th to 16th place when reducing compensation for $G_{2}$ and $G_{2}$ in conjunction with $G_{4}$ dimensions but jumps to 14th place when reducing compensation for $G_{3}$ criteria. Hungary has the lowest number of $C_{6}$ CTs and $C_{7}$ MRIs of all countries analyzed. Latvia ($A_{8}$) decreases from 14th to 16th place when reducing the compensation of criteria belonging to dimensions $G_{2}$ in conjunction with $G_{3}$ or $G_{1}$ in conjunction with $G_{2}$. Poland ($A_{12}$) is last in the ranking but advances to 14th place if compensation for $G_{2}$ criteria is reduced. Poland has the worst performance value for $C_{3}$ (practicing physicians) and $C_{25}$ (new technologies).

\subsection{Comparative analysis of obtained results with using objective weighting methods}


In the following part of the research, a comparative analysis of the results of the SSP-AHP without criteria compensation reduction for the subjective weights determined by the AHP-based relative weighting method with two other objective criteria weighting methods, that is, Entropy weighting and CRITIC weighting method was performed. This investigation was conducted to the convergency of the results provided by the compared approaches and whether the use of weights determined subjectively based on expert knowledge yields similar results as the use of weights calculated using objective methods.

\begin{figure}[H]
\centering
\includegraphics[width=0.3\linewidth]{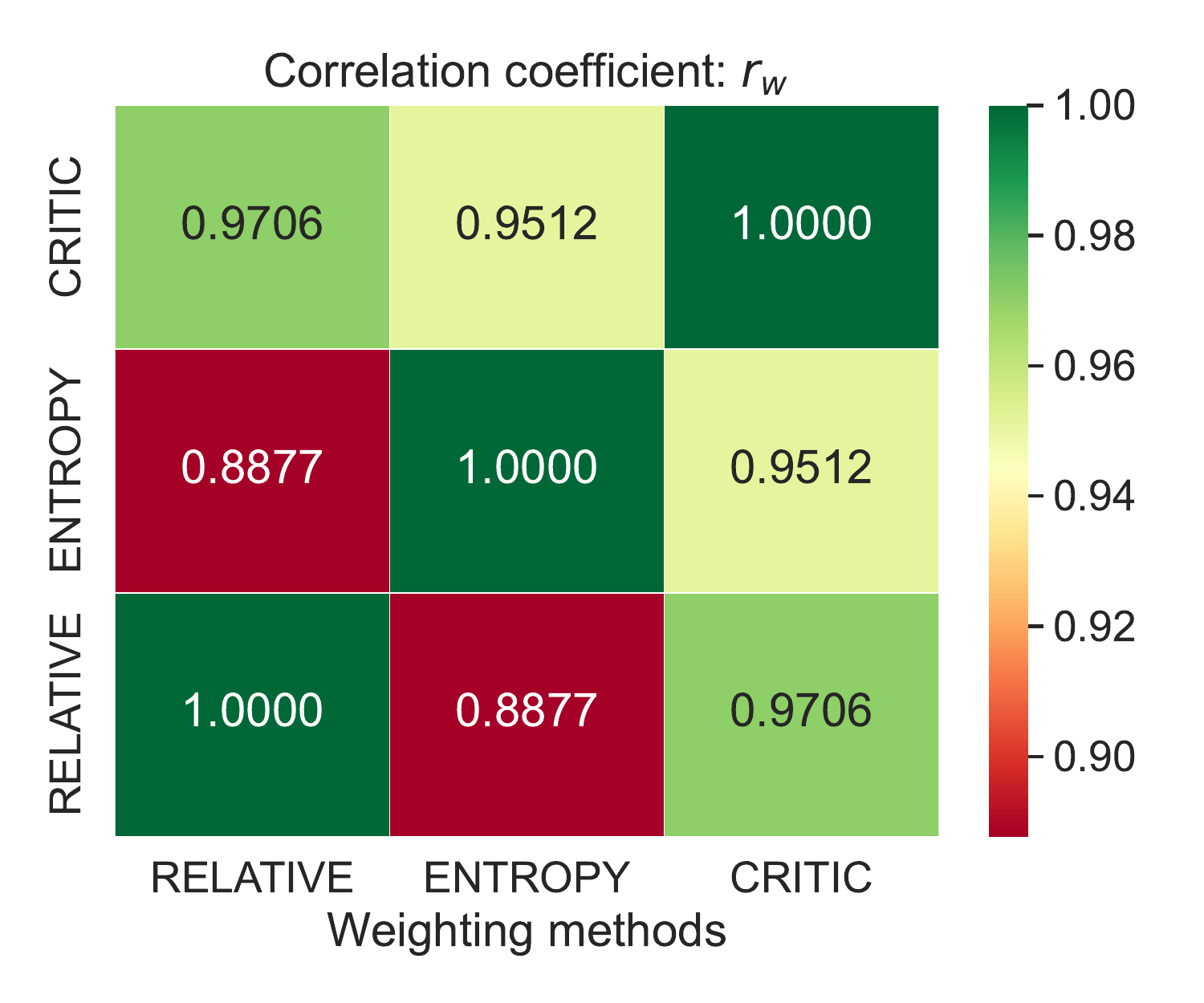}
\includegraphics[width=0.3\linewidth]{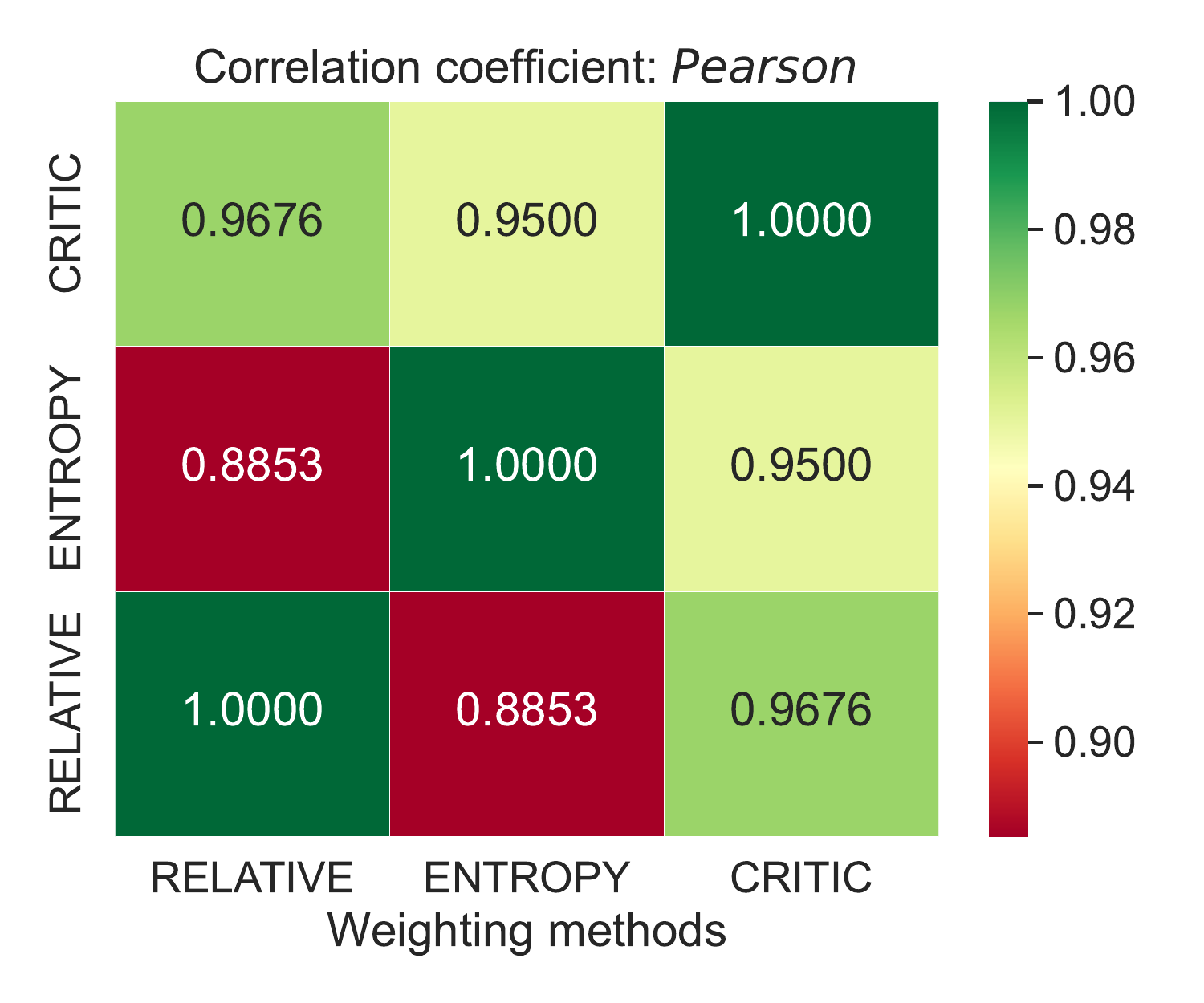}
\caption{Comparison of SSP-AHP rankings without criteria compensation reduction constructed with weights determined with different weighting methods.}
\label{fig:correlationsAHPdiffWeights}
\end{figure}

For this purpose, rankings were additionally generated using the SSP-AHP without criteria compensation reduction and benchmark methods for the criteria weights determined by objective methods. Figure~\ref{fig:correlationsAHPdiffWeights} displays the correlations determined for the rankings obtained with the SSP-AHP method without criteria compensation reduction for the criteria weights determined by the compared methods, while Table~\ref{tab:correlationsAHPdiffWeights} contains the utility function values and rankings obtained by the SSP-AHP method without criteria compensation reduction for the compared criteria weighting techniques.

From results displayed in Figure~\ref{fig:correlationsAHPdiffWeights} and contained in Table~\ref{tab:correlationsAHPdiffWeights}, it can be concluded that the criteria weights determined by the objective CRITIC method reflect the subjective preferences of the expert more accurately than the weights determined by the objective Entropy method.

\begin{figure}[H]
    \centering
    \includegraphics[width=0.7\linewidth]{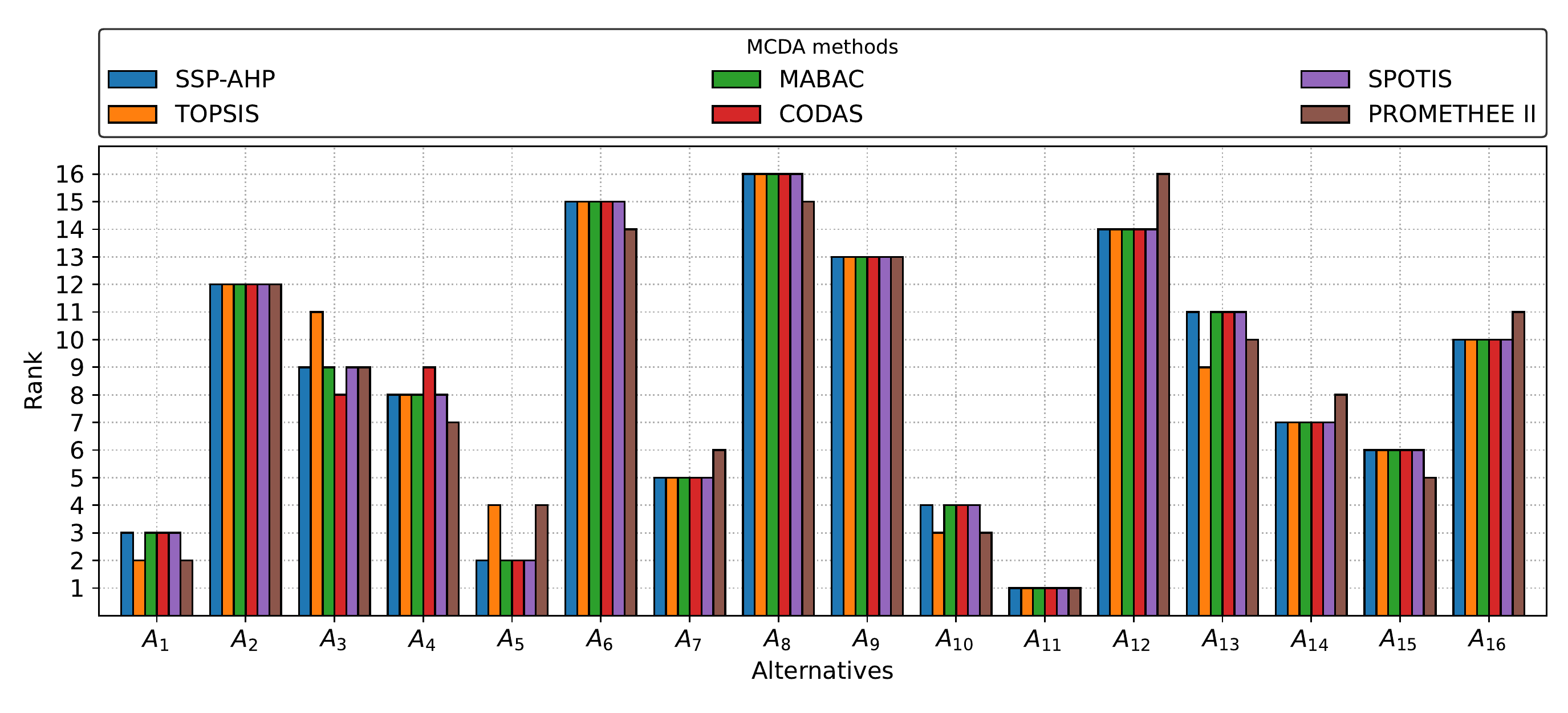}
    \caption{Comparison of rankings provided by the SSP-AHP method without criteria compensation reduction and reference MCDA methods for criteria weights determined with Entropy method.}
    \label{fig:barchartClassicEntropy}
\end{figure}

Table~\ref{tab:resultsClassicEntropy} shows the utility function values and rankings obtained using the SSP-AHP method without criteria compensation reduction for the criteria weights determined by the Entropy method. The rankings are further visualized in Figure~\ref{fig:barchartClassicEntropy}. For the criterion weights determined by the Entropy method, the correlation values of the AHP ranking with the compared rankings of the reference methods are high, as was the case with the expert weights. Similar to the use of expert criteria weights, the ranking leader is Norway ($A_{11}$). However, Sweden ($A_{15}$), in the case of Entropy weighting criteria, did not get the second-ranking position like it was noticed for the use of expert weights, but in most of the rankings, it ranked only sixth. Germany ($A_{5}$), Belgium ($A_{1}$),  Netherlands ($A_{10}$), and Iceland ($A_{7}$) were ranked high next. Thus, it can be observed that in the case of involving Entropy weighting criteria top of the six countries, the ranking includes the same alternatives, namely Norway, Sweden, Germany, Netherlands, Iceland, and Belgium.

\begin{table}[H]
\centering
\caption{The SSP-AHP results without criteria compensation reduction for criteria weights determined with different weighting methods.}
\label{tab:correlationsAHPdiffWeights}
\resizebox{\linewidth}{!}{
\begin{tabular}{llrrrrrr} \toprule
&  & \multicolumn{3}{c}{Utility function values} & \multicolumn{3}{c}{Rankings} \\ \hline
 & Country & SSP-AHP + RELATIVE weights & SSP-AHP + ENTROPY weights & SSP-AHP + CRITIC weights & SSP-AHP + RELATIVE weights & SSP-AHP + ENTROPY weights & SSP-AHP + CRITIC weights \\ \midrule
$A_{1}$ & Belgium & 0.6243 & 0.6514 & 0.6645 & 6 & 3 & 5 \\
$A_{2}$ & Czech Republic & 0.4920 & 0.4214 & 0.4789 & 11 & 12 & 12 \\
$A_{3}$ & Finland & 0.5072 & 0.4820 & 0.5654 & 9 & 9 & 8 \\
$A_{4}$ & France & 0.5776 & 0.4936 & 0.5861 & 7 & 8 & 7 \\
$A_{5}$ & Germany & 0.6775 & 0.6823 & 0.6891 & 3 & 2 & 2 \\
$A_{6}$ & Hungary & 0.3272 & 0.3178 & 0.3251 & 15 & 15 & 14 \\
$A_{7}$ & Iceland & 0.6338 & 0.5503 & 0.6553 & 5 & 5 & 6 \\
$A_{8}$ & Latvia & 0.3589 & 0.2867 & 0.3228 & 14 & 16 & 15 \\
$A_{9}$ & Luxembourg & 0.5360 & 0.3652 & 0.4847 & 8 & 13 & 11 \\
$A_{10}$ & Netherlands & 0.6391 & 0.6333 & 0.6670 & 4 & 4 & 4 \\
$A_{11}$ & Norway & 0.7311 & 0.7095 & 0.7679 & 1 & 1 & 1 \\
$A_{12}$ & Poland & 0.3132 & 0.3349 & 0.3129 & 16 & 14 & 16 \\
$A_{13}$ & Slovak Republic & 0.4001 & 0.4306 & 0.4014 & 13 & 11 & 13 \\
$A_{14}$ & Slovenia & 0.5020 & 0.4978 & 0.5553 & 10 & 7 & 9 \\
$A_{15}$ & Sweden & 0.6863 & 0.5423 & 0.6751 & 2 & 6 & 3 \\
$A_{16}$ & United Kingdom & 0.4813 & 0.4449 & 0.5100 & 12 & 10 & 10 \\ \bottomrule
\end{tabular}
}
\end{table}

\begin{figure}[H]
\centering
\includegraphics[width=0.45\linewidth]{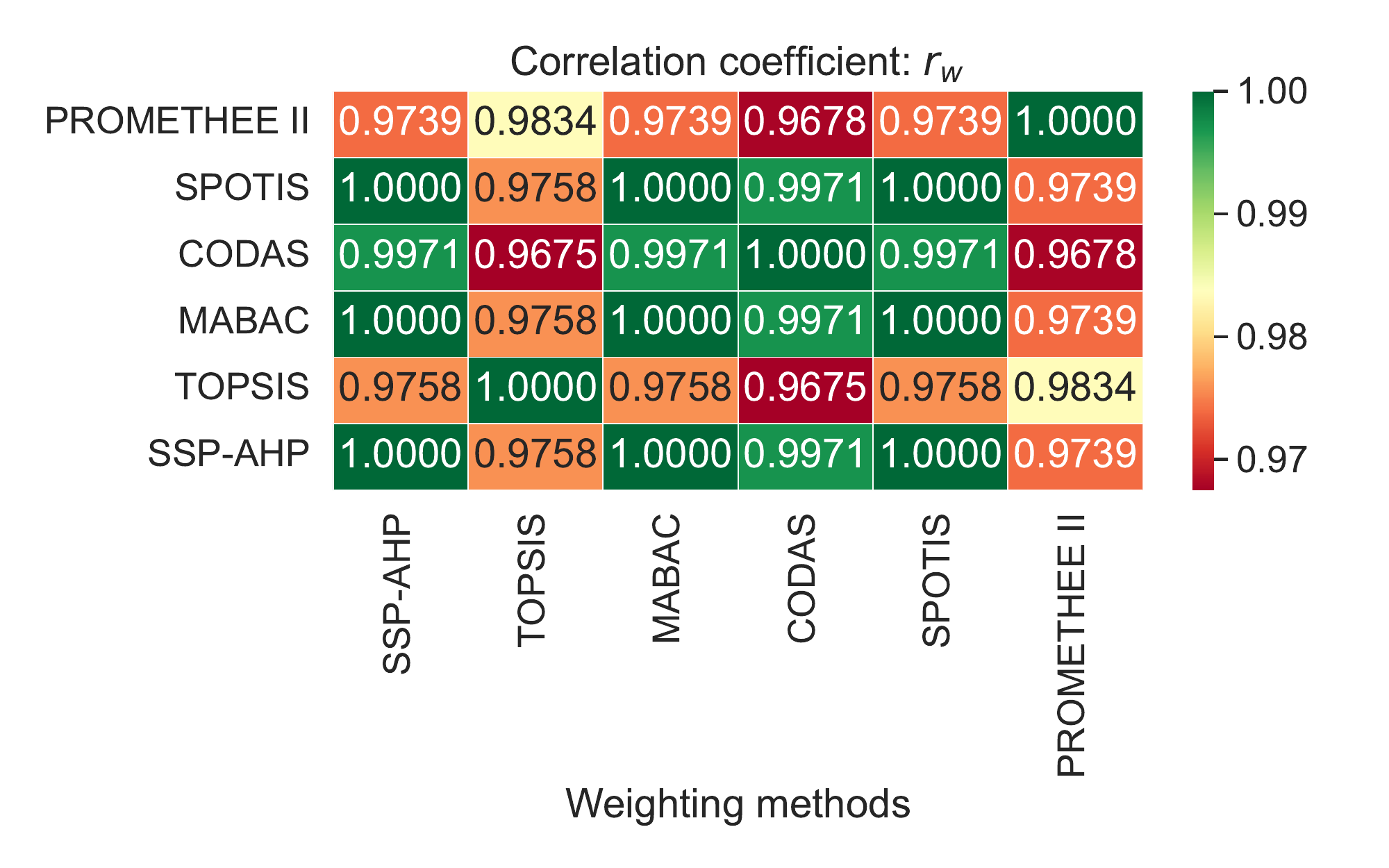}
\includegraphics[width=0.45\linewidth]{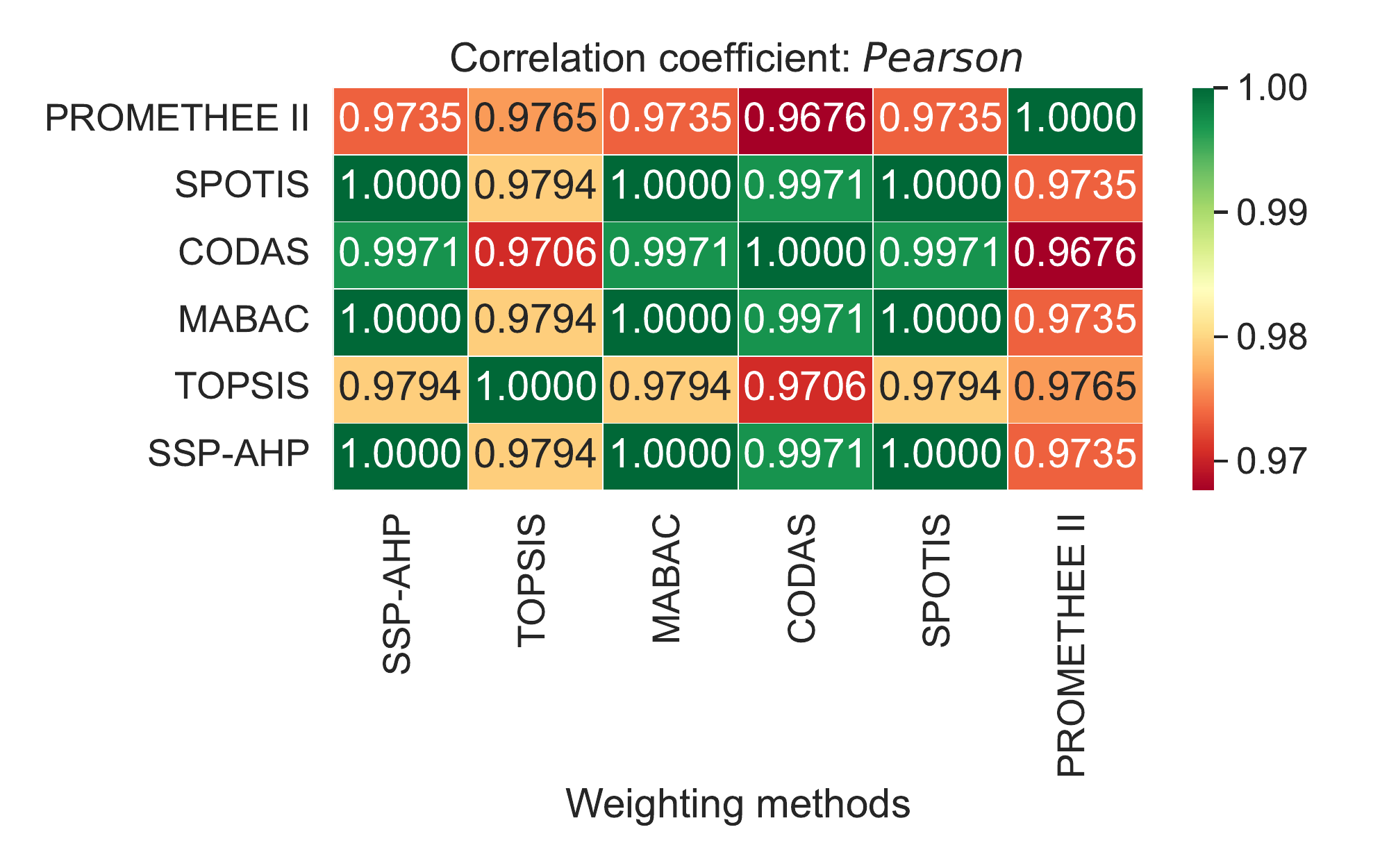}
\caption{Correlations of rankings provided by the SSP-AHP method without criteria compensation reduction for criteria weights determined with an Entropy weighting method.}
\label{fig:correlationsClassicEntropy}
\end{figure}

\begin{table}[H]
\centering
\caption{Utility function values and rankings provided by the SSP-AHP method without criteria compensation reduction and reference methods using weights determined with an objective Entropy weighting method.}
\label{tab:resultsClassicEntropy}
\resizebox{\linewidth}{!}{
\begin{tabular}{llrrrrrrrrrrrr} \toprule
$A_{i}$ & Country & SSP-AHP & TOPSIS & MABAC & CODAS & SPOTIS & PROMETHEE II & SSP-AHP & TOPSIS & MABAC & CODAS & SPOTIS & PROMETHEE II \\ \midrule
$A_{1}$ & Belgium & 0.6514 & 0.6415 & 0.1870 & 3.3776 & 0.3486 & 0.3521 & 3 & 2 & 3 & 3 & 3 & 2 \\
$A_{2}$ & Czech Republic & 0.4214 & 0.4416 & -0.0429 & -1.0090 & 0.5786 & -0.1943 & 12 & 12 & 12 & 12 & 12 & 12 \\
$A_{3}$ & Finland & 0.4820 & 0.4607 & 0.0177 & -0.1666 & 0.5180 & -0.0431 & 9 & 11 & 9 & 8 & 9 & 9 \\
$A_{4}$ & France & 0.4936 & 0.4862 & 0.0292 & -0.3927 & 0.5064 & 0.0660 & 8 & 8 & 8 & 9 & 8 & 7 \\
$A_{5}$ & Germany & 0.6823 & 0.6322 & 0.2179 & 3.6040 & 0.3177 & 0.3406 & 2 & 4 & 2 & 2 & 2 & 4 \\
$A_{6}$ & Hungary & 0.3178 & 0.3561 & -0.1466 & -3.2870 & 0.6822 & -0.3389 & 15 & 15 & 15 & 15 & 15 & 14 \\
$A_{7}$ & Iceland & 0.5503 & 0.5143 & 0.0859 & 0.8097 & 0.4497 & 0.1037 & 5 & 5 & 5 & 5 & 5 & 6 \\
$A_{8}$ & Latvia & 0.2867 & 0.3183 & -0.1776 & -3.7885 & 0.7133 & -0.3680 & 16 & 16 & 16 & 16 & 16 & 15 \\
$A_{9}$ & Luxembourg & 0.3652 & 0.3898 & -0.0992 & -2.4052 & 0.6348 & -0.2516 & 13 & 13 & 13 & 13 & 13 & 13 \\
$A_{10}$ & Netherlands & 0.6333 & 0.6344 & 0.1689 & 2.9944 & 0.3667 & 0.3475 & 4 & 3 & 4 & 4 & 4 & 3 \\
$A_{11}$ & Norway & 0.7095 & 0.6572 & 0.2451 & 4.1099 & 0.2905 & 0.5086 & 1 & 1 & 1 & 1 & 1 & 1 \\
$A_{12}$ & Poland & 0.3349 & 0.3821 & -0.1294 & -3.0782 & 0.6651 & -0.3695 & 14 & 14 & 14 & 14 & 14 & 16 \\
$A_{13}$ & Slovak Republic & 0.4306 & 0.4714 & -0.0338 & -0.5317 & 0.5694 & -0.1292 & 11 & 9 & 11 & 11 & 11 & 10 \\
$A_{14}$ & Slovenia & 0.4978 & 0.4922 & 0.0334 & -0.0458 & 0.5022 & -0.0301 & 7 & 7 & 7 & 7 & 7 & 8 \\
$A_{15}$ & Sweden & 0.5423 & 0.4922 & 0.0779 & 0.3248 & 0.4577 & 0.2000 & 6 & 6 & 6 & 6 & 6 & 5 \\
$A_{16}$ & United Kingdom & 0.4449 & 0.4679 & -0.0195 & -0.5155 & 0.5551 & -0.1936 & 10 & 10 & 10 & 10 & 10 & 11 \\ \bottomrule
\end{tabular}
}
\end{table}
%
The described similarity of results for the subjective and objective criteria weights is confirmed by the high correlation coefficients values of the rankings displayed in Figure~\ref{fig:correlationsClassicEntropy}.

\begin{figure}[H]
    \centering
    \includegraphics[width=0.7\linewidth]{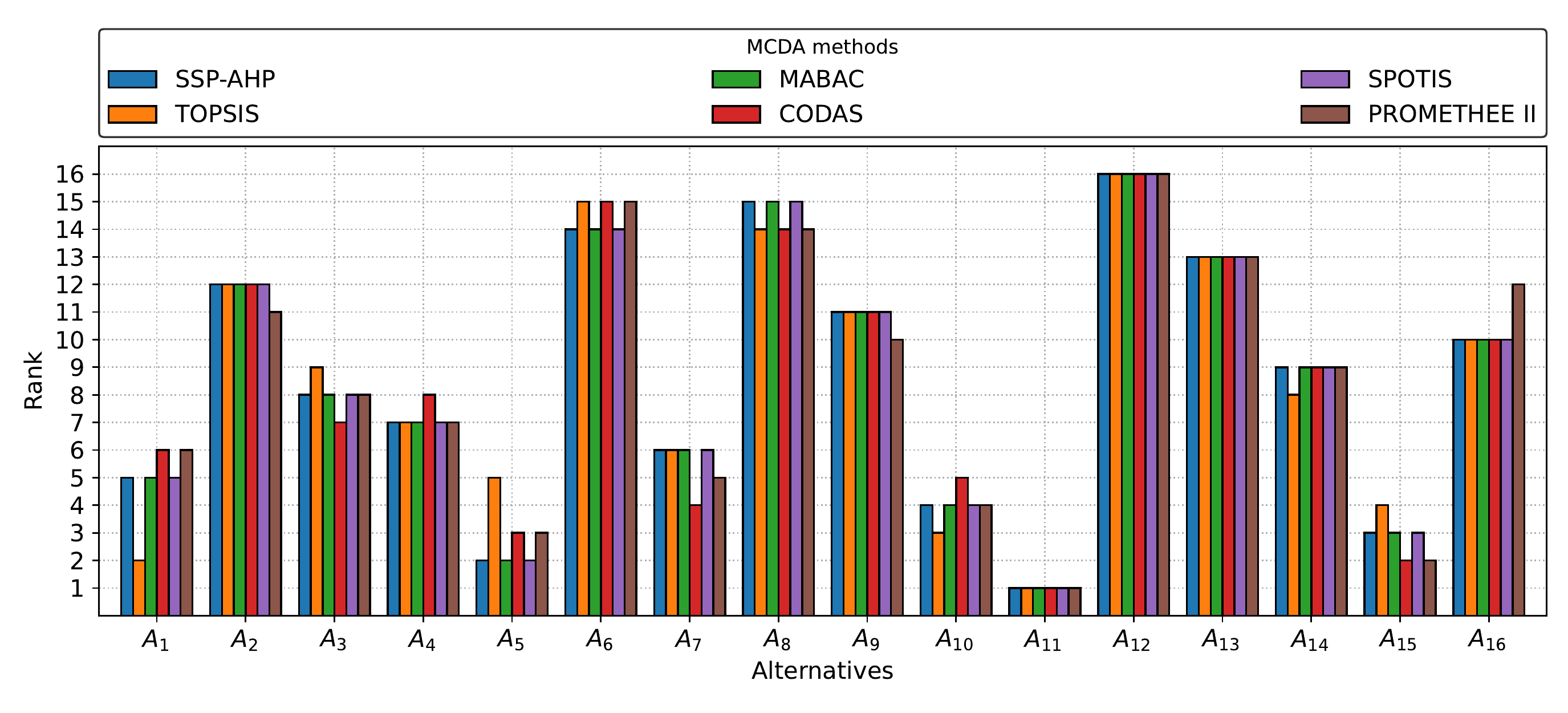}
    \caption{Comparison of rankings provided by the SSP-AHP method without criteria compensation reduction and reference MCDA methods for criteria weights determined with CRITIC method.}
    \label{fig:barchartClassicCRITIC}
\end{figure}

Table~\ref{tab:resultsClassicCRITIC} contains the utility function values and rankings obtained applying the SSP-AHP method without criteria compensation reduction and the reference methods using objective weights calculated by the CRITIC method.
\begin{table}[H]
\centering
\caption{Utility function values and rankings provided by the SSP-AHP method without criteria compensation reduction and reference methods using weights determined with an objective CRITIC weighting method.}
\label{tab:resultsClassicCRITIC}
\resizebox{\linewidth}{!}{
\begin{tabular}{llrrrrrrrrrrrr} \toprule
$A_{i}$ & Country & SSP-AHP & TOPSIS & MABAC & CODAS & SPOTIS & PROMETHEE II & SSP-AHP & TOPSIS & MABAC & CODAS & SPOTIS & PROMETHEE II \\ \midrule
$A_{1}$ & Belgium & 0.6645 & 0.6427 & 0.1506 & 2.0106 & 0.3355 & 0.1992 & 5 & 2 & 5 & 6 & 5 & 6 \\
$A_{2}$ & Czech Republic & 0.4789 & 0.4898 & -0.0350 & -1.1498 & 0.5211 & -0.1698 & 12 & 12 & 12 & 12 & 12 & 11 \\
$A_{3}$ & Finland & 0.5654 & 0.5510 & 0.0515 & 0.7856 & 0.4346 & 0.0219 & 8 & 9 & 8 & 7 & 8 & 8 \\
$A_{4}$ & France & 0.5861 & 0.5744 & 0.0722 & 0.7154 & 0.4139 & 0.0719 & 7 & 7 & 7 & 8 & 7 & 7 \\
$A_{5}$ & Germany & 0.6891 & 0.6307 & 0.1752 & 2.4737 & 0.3109 & 0.2962 & 2 & 5 & 2 & 3 & 2 & 3 \\
$A_{6}$ & Hungary & 0.3251 & 0.3669 & -0.1888 & -4.0349 & 0.6749 & -0.3616 & 14 & 15 & 14 & 15 & 14 & 15 \\
$A_{7}$ & Iceland & 0.6553 & 0.6143 & 0.1414 & 2.3248 & 0.3447 & 0.2238 & 6 & 6 & 6 & 4 & 6 & 5 \\
$A_{8}$ & Latvia & 0.3228 & 0.3752 & -0.1911 & -3.7687 & 0.6772 & -0.3467 & 15 & 14 & 15 & 14 & 15 & 14 \\
$A_{9}$ & Luxembourg & 0.4847 & 0.4952 & -0.0292 & -0.6698 & 0.5153 & -0.1110 & 11 & 11 & 11 & 11 & 11 & 10 \\
$A_{10}$ & Netherlands & 0.6670 & 0.6417 & 0.1531 & 2.0438 & 0.3330 & 0.2676 & 4 & 3 & 4 & 5 & 4 & 4 \\
$A_{11}$ & Norway & 0.7679 & 0.7074 & 0.2540 & 3.9955 & 0.2321 & 0.5310 & 1 & 1 & 1 & 1 & 1 & 1 \\
$A_{12}$ & Poland & 0.3129 & 0.3535 & -0.2010 & -4.3977 & 0.6871 & -0.4843 & 16 & 16 & 16 & 16 & 16 & 16 \\
$A_{13}$ & Slovak Republic & 0.4014 & 0.4323 & -0.1125 & -2.5637 & 0.5986 & -0.2906 & 13 & 13 & 13 & 13 & 13 & 13 \\
$A_{14}$ & Slovenia & 0.5553 & 0.5623 & 0.0414 & 0.3020 & 0.4447 & -0.0166 & 9 & 8 & 9 & 9 & 9 & 9 \\
$A_{15}$ & Sweden & 0.6751 & 0.6320 & 0.1612 & 2.4964 & 0.3249 & 0.3774 & 3 & 4 & 3 & 2 & 3 & 2 \\
$A_{16}$ & United Kingdom & 0.5100 & 0.5199 & -0.0039 & -0.5631 & 0.4900 & -0.2086 & 10 & 10 & 10 & 10 & 10 & 12 \\ \bottomrule
\end{tabular}
}
\end{table}
In addition, the obtained rankings are visualized in the column graph shown in Figure~\ref{fig:barchartClassicCRITIC}. 

When using the CRITIC method weights, the multi-criteria evaluation with the classic AHP and reference methods identified Norway as the leader in the sustainable healthcare system, with Germany, Sweden, Netherlands, Belgium, and Iceland among the best-scored countries. However, a swap of one position was observed for Sweden and Germany and Iceland and Belgium compared with ranking provided with subjective weights.

\begin{figure*}[ht!]
\centering
\includegraphics[width=0.45\linewidth]{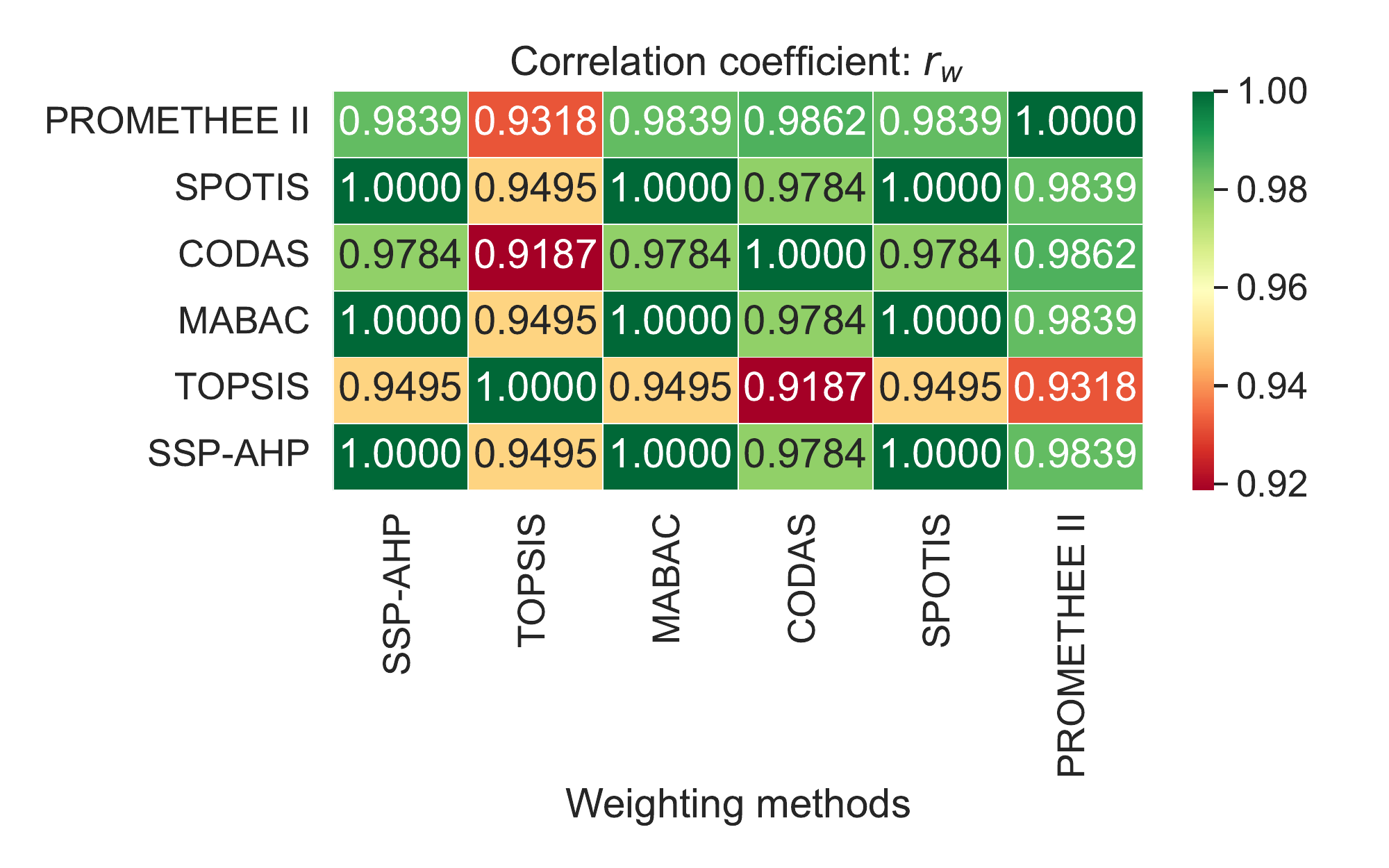}
\includegraphics[width=0.45\linewidth]{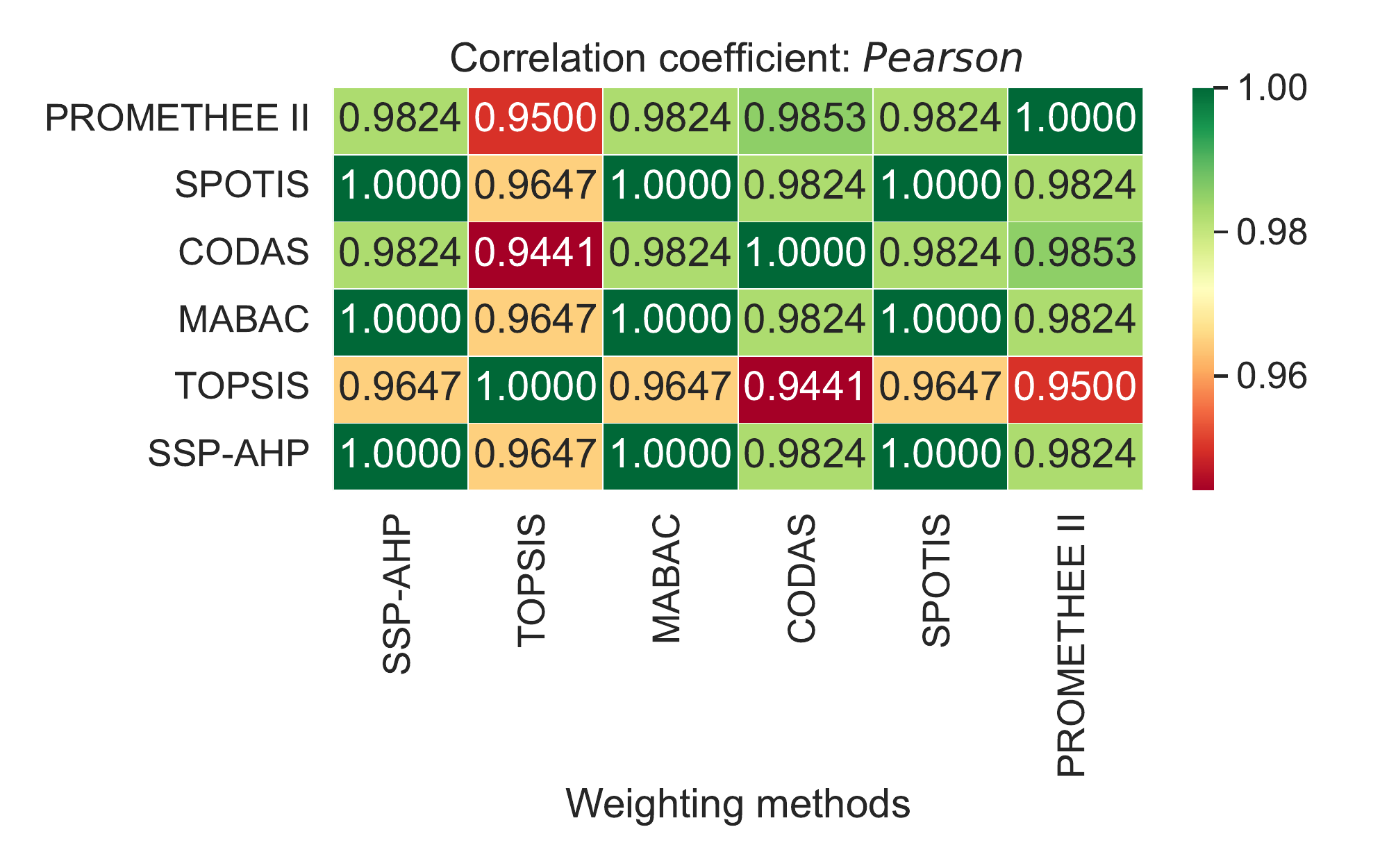}
\caption{Correlations of rankings provided by the SSP-AHP method without criteria compensation reduction and reference methods for criteria weights determined with CRITIC weighting method.}
\label{fig:correlationsClassicCRITIC}
\end{figure*}

The values of the correlation coefficients calculated for the rankings provided by the SSP-AHP method without criteria compensation reduction and reference methods for the weights determined by the CRITIC method, visualized in Figure~\ref{fig:correlationsClassicCRITIC}, are high, as in the case of the AHP-based relative and Entropy weighting methods.
%
%

\begin{figure}[H]
    \centering
    \includegraphics[width=0.5\linewidth]{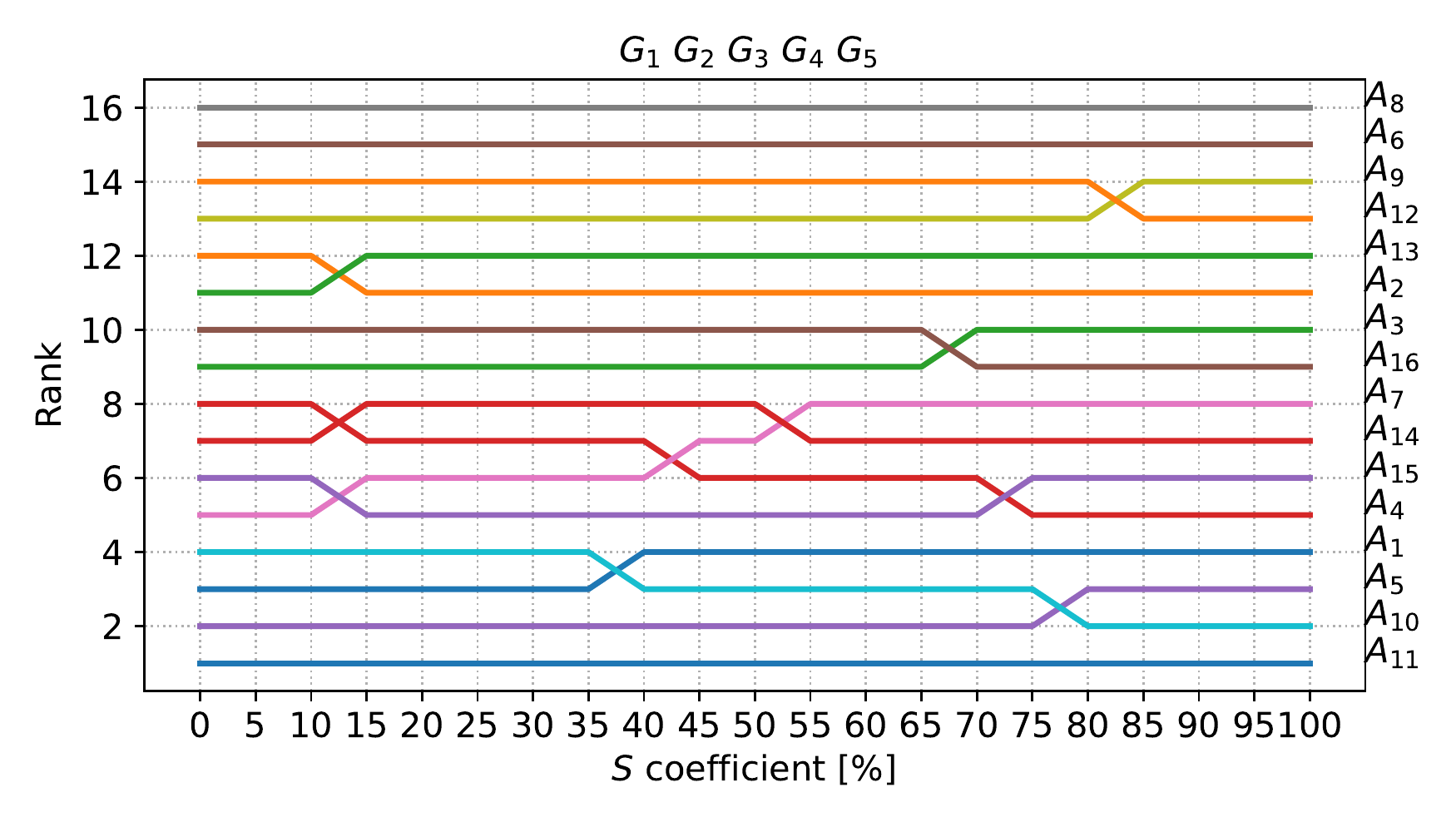}
    \caption{Changes in SSP-AHP ranking with modification of $s$ coefficient in all criteria for criteria weights determined with Entropy method.}
    \label{fig:sustCoeffEntropy}
\end{figure}

The next phase of the research aimed to compare rankings obtained by the SSP-AHP method, using criteria weights determined subjectively by the expert with the AHP method based on the Saaty scale with SSP-AHP rankings obtained for weights determined using objective Entropy weighting and CRITIC weighting methods. The changes in rankings as a result of reducing the compensation of all criteria by gradually increasing the coefficient $s$ are visualized in Figure~\ref{fig:sustCoeffEntropy}.

It is worth noting that Germany ($A_{5}$) drops from second to fifth place when the compensation of the main conjoined criteria $G_{1}$, $G_{4}$, and $G_{5}$ is reduced but moves up to a first position when the compensation of criteria belonging to $G_{3}$ is reduced. Variability with a reduction in criteria compensation is also shown by Sweden ($A_{15}$) advancing from sixth place to fourth place when compensation of conjoined criteria $G_{1}$, $G_{4}$, and $G_{4}$ is reduced and falling to eighth place when compensation of conjoined criteria $G_{2}$, $G_{3}$, and $G_{4}$ is reduced. In contrast, France ($A_{4}$) tends to climb from eighth place to fifth place. The best-ranked and most stable countries are the Netherlands ($A_{10}$), which jumps from fourth to second place when criteria compensation is reduced, and Norway ($A_{11}$), the stable leader of the ranking. On the other hand, the lowest-ranked countries for which no significant change was observed in the reduction of criteria compensation in the SSP-AHP method are Latvia ($A_{8}$), which is stable in last place, Hungary ($A_{6}$), Poland ($A_{12}$), Slovak Republic ($A_{13}$). 

%

For the analogous investigation using CRITIC-determined weights, the highest variability when reducing criterion compensation with increasing $s$ is also observed for Germany ($A_{5}$). Germany drops from second to eighth place when reducing the compensation of conjoined criteria $G_{1}$, $G_{3}$, $G_{4}$, $G_{5}$. Netherlands ($A_{10}$) and France ($A_{4}$) tend to move up when reducing criteria compensation.

\begin{figure}[H]
    \centering
    \includegraphics[width=0.5\linewidth]{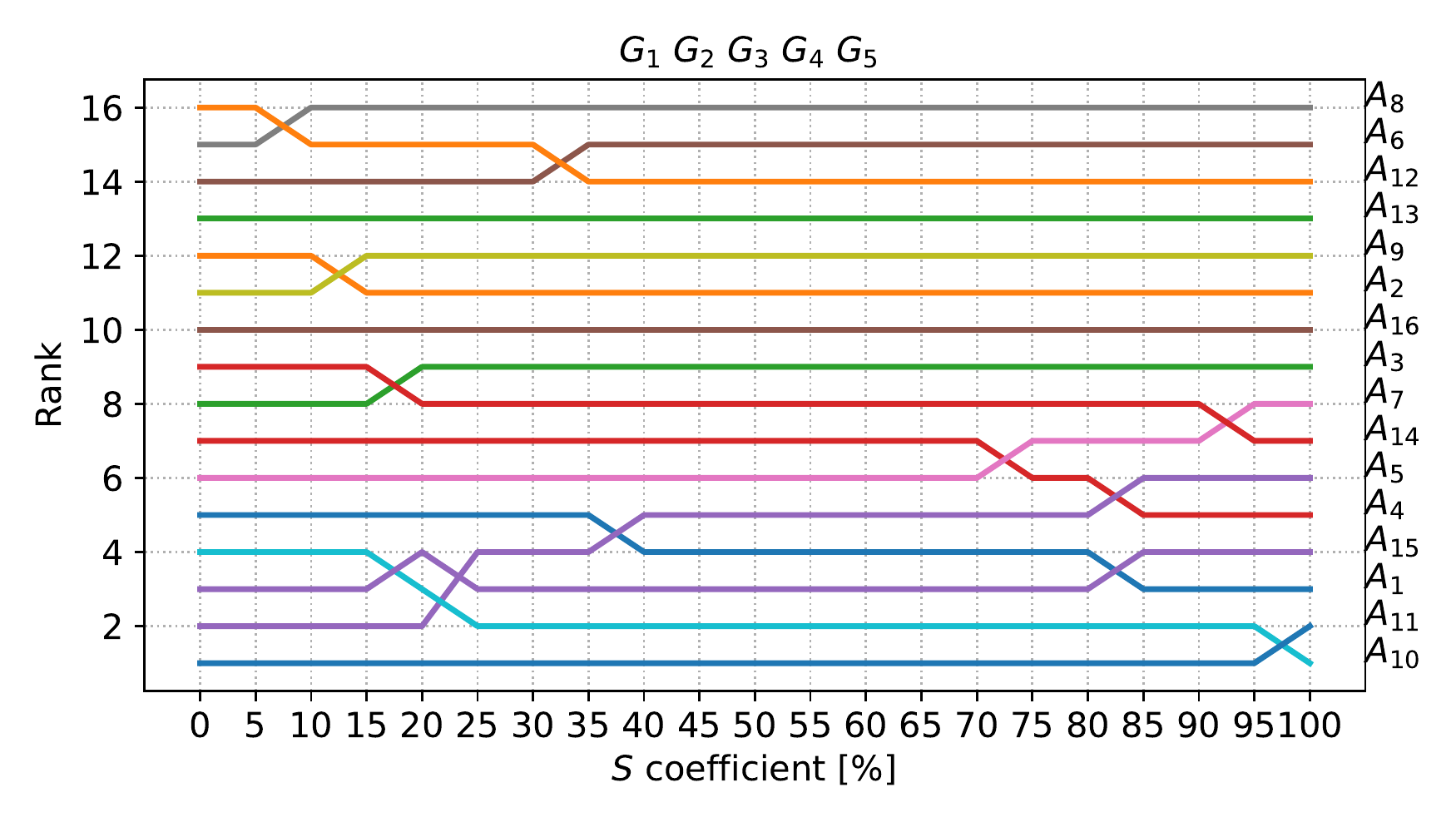}
    \caption{Changes in SSP-AHP ranking with modification of $s$ coefficient in all criteria for criteria weights determined with CRITIC method.}
    \label{fig:sustCoeffCRITIC}
\end{figure}

Norway ($A_{11}$) is a stable leader, falling only in the case of maximum compensation reduction for all criteria to second place. Sweden ($A_{15}$) drops from third to fifth place when reducing the compensation of criteria belonging to $G_{2}$ but moves up to second place when reducing the compensation of conjoined criteria $G_{5}$, $G_{4}$ and $G_{5}$, $G_{3}$ in conjunction with $G_{5}$, $G_{3}$ in conjunction with $G_{4}$. In the last, stable places are Latvia ($A_{8}$), Poland ($A_{12}$), and Slovak Republic ($A_{13}$). The changes in rankings as a result of reducing the compensation of all criteria by gradually increasing the coefficient $s$ are visualized in Figure~\ref{fig:sustCoeffCRITIC}.
%
\begin{table}[H]
\centering
\caption{Correlation coefficient values of SSP-AHP rankings obtained for subjective weights applying criteria compensation with SSP-AHP rankings obtained using weights determined by Entropy and CRITIC objective methods.}
\label{tab:corrWeightingMethods}
\resizebox{\linewidth}{!}{
\begin{tabular}{lrrrr}
\toprule
$G$ & $r_w$ for SSP-AHP + CRITIC weights & Pearson for SSP-AHP + CRITIC weights & $r_w$ for SSP-AHP + Entropy weights & Pearson for SSP-AHP + Entropy weights \\ \midrule
 & 0.9706 & 0.9676 & 0.8877 & 0.8853 \\
$G_{5}$ & 0.9685 & 0.9618 & 0.9100 & 0.9029 \\
$G_{4}$ & 0.9403 & 0.9441 & 0.8593 & 0.8706 \\
$G_{4}$ $G_{5}$ & 0.9685 & 0.9618 & 0.8760 & 0.8853 \\
$G_{3}$ & 0.9696 & 0.9618 & 0.9035 & 0.8882 \\
$G_{3}$ $G_{5}$ & 0.9645 & 0.9588 & 0.9031 & 0.8765 \\
$G_{3}$ $G_{4}$ & 0.9524 & 0.9500 & 0.8936 & 0.9000 \\
$G_{3}$ $G_{4}$ $G_{5}$ & 0.9569 & 0.9529 & 0.8818 & 0.8647 \\
$G_{2}$ & 0.9550 & 0.9500 & 0.9455 & 0.9353 \\
$G_{2}$ $G_{5}$ & 0.9550 & 0.9500 & 0.9320 & 0.9147 \\
$G_{2}$ $G_{4}$ & 0.9438 & 0.9412 & 0.9450 & 0.9324 \\
$G_{2}$ $G_{4}$ $G_{5}$ & 0.9388 & 0.9353 & 0.9391 & 0.9176 \\
$G_{2}$ $G_{3}$ & 0.9559 & 0.9559 & 0.9144 & 0.9118 \\
$G_{2}$ $G_{3}$ $G_{5}$ & 0.9683 & 0.9647 & 0.9292 & 0.9235 \\
$G_{2}$ $G_{3}$ $G_{4}$ & 0.9521 & 0.9529 & 0.9140 & 0.9118 \\
$G_{2}$ $G_{3}$ $G_{4}$ $G_{5}$ & 0.9471 & 0.9500 & 0.9360 & 0.9294 \\
$G_{1}$ & 0.9574 & 0.9618 & 0.8689 & 0.8794 \\
$G_{1}$ $G_{5}$ & 0.9526 & 0.9500 & 0.8740 & 0.8794 \\
$G_{1}$ $G_{4}$ & 0.9375 & 0.9441 & 0.8638 & 0.8735 \\
$G_{1}$ $G_{4}$ $G_{5}$ & 0.9526 & 0.9500 & 0.8964 & 0.8971 \\
$G_{1}$ $G_{3}$ & 0.9410 & 0.9441 & 0.8609 & 0.8735 \\
$G_{1}$ $G_{3}$ $G_{5}$ & 0.9450 & 0.9441 & 0.8817 & 0.8706 \\
$G_{1}$ $G_{3}$ $G_{4}$ & 0.9353 & 0.9382 & 0.8348 & 0.8529 \\
$G_{1}$ $G_{3}$ $G_{4}$ $G_{5}$ & 0.9426 & 0.9412 & 0.8296 & 0.8500 \\
$G_{1}$ $G_{2}$ & 0.9415 & 0.9412 & 0.9170 & 0.9176 \\
$G_{1}$ $G_{2}$ $G_{5}$ & 0.9503 & 0.9471 & 0.9170 & 0.9176 \\
$G_{1}$ $G_{2}$ $G_{4}$ & 0.9415 & 0.9412 & 0.9045 & 0.9059 \\
$G_{1}$ $G_{2}$ $G_{4}$ $G_{5}$ & 0.9637 & 0.9588 & 0.9102 & 0.9118 \\
$G_{1}$ $G_{2}$ $G_{3}$ & 0.9472 & 0.9500 & 0.8938 & 0.9118 \\
$G_{1}$ $G_{2}$ $G_{3}$ $G_{5}$ & 0.9649 & 0.9618 & 0.9260 & 0.9206 \\
$G_{1}$ $G_{2}$ $G_{3}$ $G_{4}$ & 0.9495 & 0.9529 & 0.8926 & 0.9088 \\
$G_{1}$ $G_{2}$ $G_{3}$ $G_{4}$   $G_{5}$ & 0.9559 & 0.9559 & 0.8984 & 0.9029 \\ \bottomrule
\end{tabular}
}
\end{table}

The results obtained using the two objective criteria weighting methods are not identical to the results for the subjectively determined weights, but they reflect the locations and trends obtained for the subjective weights.
The final step was to objectively determine the convergence of the rankings provided by the SSP-AHP method using a subjective criterion weighting method for reducing the compensation of individual combinations of the main $G_{1}$--$G_{5}$ criterion groups compared to the corresponding results obtained using weights calculated with the Entropy and CRITIC methods. In order to ensure the reliability of results, correlations were determined using two coefficients: $r_w$ and Pearson correlation coefficient. SSP-AHP using AHP weighting method correlation values compared to employing Entropy and CRITIC weighting are included in Table~\ref{tab:corrWeightingMethods}. Similar to the SSP-AHP method without criteria compensation reduction, for the SSP-AHP method with reduction of criteria compensation, a higher correlation of rankings obtained using subjective weights was observed for comparisons with rankings obtained for the CRITIC weighting method. Besides, the high values of all obtained measurements confirm the reliability of the expert weights used in the research and the SSP-AHP method.
\section{Discussion}
\label{sec:discussion}

Our understanding of social sustainability shares the perspective proposed by Rachelle et al.~\citep{hollander2016network}, highlighting the quality of society that encourages durable environment for human well-being. The present paper is among the first studies to map social sustainability values in health systems from the national level. The proposed framework offers a comprehensive assessment of  social sustainable approach of health systems and serves as an analytical instrument for cross-national comparisons. It has been based on five major domains: equity, quality, responsiveness, financial coverage and adaptability. The first four of these are in line with the broadly recognized health system performance frameworks, introduced by WHO and OECD~\citep{murray2000framework}. Although the proposed framework covers the most crucial, well-known dimensions, reflecting the principal aims of health system, it also captures up-to-date dimension i.e. adaptability. It seems extremely important to boost resilience and flexibility of health systems, especially in times of crisis, such as COVID pandemic. This opinion is shared  among others by Macassa and Tomaselli~\citep{macassa2020rethinking}.

The proposed framework has some potential for scientific contribution, embedding novel approach towards sustainability assessment methodology. It also enables cross-national comparisons and thus can be applied as an effective instrument measuring the level of social sustainability of health systems. Thus, it has practical implications for health policy leaders and policy makers. The innovative value of the study is also based on dynamic approach towards assessing social sustainability phenomenon, which remains in contract to static frameworks introduced by WHO and OECD. The dynamism of our approach involves the flexibility and substitutability of core five domains. The previous studies concentrated on the organizational perspective, focusing on themes related to corporate management~\citep{aljaberi2020framework, cristiano2021systemic, jahani2021sustainability} and supply chain management~\citep{hussain2019exploration} and its impact on social sustainability. Another work by Ajmal et al.~\citep{ajmal2018conceptualizing} adopted both external, societal prism and internal, companies’ perspective, introducing six major social sustainability areas. The areas related to external perspective, namely social development, social growth and social justice can be compared to our core domains: financial coverage, quality (in terms of health outcomes), and equity (in terms of accessibility). 

The findings reveal that the most sustainable health systems, no matter which MCDM method has been applied, are those representing so called the Nordic model. It contains consistent features, such as tax-based funding, universal access on the basis of residency and comprehensive coverage of health services~\citep{magnussen2009ebook}. The Nordic countries uphold the principles of equity and universalism, which are deeply rooted in the welfare tradition. From the axiological point of view, social sustainability is thus underpinned in fundamental social values of freedom and equality~\citep{garces2003towards}. Our findings support the thesis that in the Nordic countries universal access to high-quality healthcare is a priority goal of the health system, shared commonly across the political parties~\citep{magnussen2009ebook}. Health policies in Nordic countries are largely guided by egalitarian rule, which proves to be effective in attaining improvements in health outcomes. The responsibility for providing healthcare to the population is decentralized and lies at the regional level. The stable position of Nordic health systems in terms of social sustainability provides clear evidence that decentralized distribution of resources, their needs-based allocation and regional propriety setting result in effective strategies, reducing inequalities in health~\citep{arnesen2002gender, lahelma2002analysing}. Decentralization of resource allocation is also a mean to improve responsiveness through the “short distance” between patients and local authorities, responsible for providing care. All Nordic countries have introduced measures to strengthen the role of patients~\citep{andersson2007does} and to involve them in decision making regarding treatment.

Our empirical evidence is in line with the findings of other authors~\citep{carey2015towards, greve2016migrants}, supporting the thesis that the Nordic priority setting models safeguard equity, remaining effective in offering high-quality and responsive medical services. On the other hand, the knowledge-based solutions underpinning the measurement of outcomes of different medical interventions support systems’ accountability. The Nordic countries are actively involved in technology assessment movements and quality programs that give grater scope for results. Life expectancy in Norway, the leader of the social sustainable health system ranking, is the highest in Europe, at 83.3 years in 2020 and still increasing (by 0.3 years in 2020), despite the COVID-19 pandemic~\citep{norway2021}. The level of system’s adaptability in Nordic countries is also high, resulting in the strong resilience. The perfect illustration is COVID crisis, and the relative success of handling it, attributed to factors such as rapidly implemented containment measurements, high level of trust in government and broad social welfare arrangements~\citep{norway2021}. This matches the research results obtained by applying novel SSP-AHP method.

The study findings suggest that there is a group of health systems, that manifests the most significant susceptibility to a gradual reduction in criteria of social sustainability. Among these countries are Germany, France, Netherlands, Belgium and Slovenia. However, this group is not homogeneous. The greatest shift in the ranking has been observed in case of France, and the biggest drop in case of Germany. The variations in above mentioned susceptibility can be explained on the basis of differences in the functioning of these systems, although there are certain common aspects, such as social insurance based on the Bismarck’s model. French and Slovenian health systems have a more centralized character with the more powerful position for the state than German and Dutch statutory health insurance (SHI). Another crucial difference is that the French SHI has not had the management responsibilities assigned to SHI funds and regimes, such as those observed in Germany~\citep{chevreul2015france}. In France financial and operational management is located mainly at the state level. French system receives one of the highest public financing of healthcare expenditure in Europe (11,1\% of GDP in 2019, above the EU average of 9,9\%), while out-of-pocket spending is among the lowest (9,3\% compared to 15,4\% EU average)~\citep{france2021}. Our investigation suggests that French system is relatively stable in all dimensions and the most significant advancement has been noted while decreasing compensation for all criteria except adaptability. This is in line with the previous findings, supporting the discussion on the significant role of voluntary health insurance (VHI) in ensuring equity in access and financial risk protection. French VHI provides complementary insurance, for instance co-payments and better coverage for medical goods and services that are poorly covered by SHI. It finances 13.8\% of total health expenditure and covers more than 90\% of the population~\citep{chevreul2015france}. Relative sound responsiveness of the system results in very low level of unmet needs for medical care, reported by only 1,2\% of the population in 2019~\citep{france2021}. These numbers support the empirical evidence based on SSP-AHP. 

Furthermore, according to our findings German heath system demonstrates the most significant susceptibility to a gradual reduction in criteria compensation such as equity in access and financial coverage. This result can be attributed to the broad benefits package within universal SHI, offered by German health system. Works by other authors support these findings, highlighting that Germany scores very well in accessibility to medical services~\citep{osborn2016new}. Public surveys reveal at the same time, that unmet needs for healthcare are driven less by financial reasons than by other factors such as waiting times. The share of patients reporting cost burden for a medical examination was only 0.1\% in 2019~\citep{blumel2021germany}.

Finally we have observed that Poland, Latvia and Hungary compose the group of countries that score very low in all dimensions. Their health systems lack social sustainability. Despite some differences in financing and provision of care (single-payer health insurance system in Poland and in Hungary; general taxation in Latvia), they all face common, ongoing challenges, affecting social sustainability. Although statutory health cover is, by definition, universal there, inequalities in care use, long waiting times for certain services and a high proportion of population with unmet medical needs let us to conclude that the “universality” formula is not assured. Similar remarks have been presented by other researchers~\citep{behmane2019latvia, gaal2011hungary, sowada2019poland}. Current health expenditures (as shares of GDP, 2019) are much below EU average (6,4\% in Hungary, 6,5\% in Poland, 6,6\% in Latvia, compared to 9,9\% across the UE~\citep{oecd2021stat}) and out-of-pocket spending (excluding informal payments) still accounts for over a third of total. Moreover, the high amenable mortality rate and the low cancer survival rates are further indicators of deficiencies in the quality of curative care~\citep{hungary2021, latvia2021, poland2021}. All discussed countries experience persistent workforce shortages that lead to low systems’ responsiveness. It can be concluded that a key challenge in ensuring sustainable health systems in central-eastern Europe is to increase the share of public expenditure on health and reduce the substantial dependence on out-of-pocket payments.

\section{Conclusions}
\label{sec:conclusions}
Health system performance assessment is a fundamental instrument for evidence-based governance in health sectors~\citep{blumel2020integrating}. A lot of effort has been put so far into the development of frameworks measuring the accountability of health systems. The publication of "Our common future''~\citep{wced1987world} in 1987 triggered a rethinking on environment, development, and governance, also in the health sector. In this perspective, the design of a sustainable performance evaluation framework seems to be fundamental to aligning health systems' operation with fulfilling the sustainability goals (SG) of Agenda 2030, especially SG 3. With this work, authors seek to map social sustainability values in health systems from the national level. The multi-stage model links the classical AHP method with the new approach called SSP-AHP into a unique methodological framework that allows prioritization of the identified values in social-sustainable health systems. The proposed conceptual and structured framework has been built upon five pillars (equity, quality, responsiveness, financial coverage, and adaptability) and 25 indicators that can guide the measurement of the social sustainability-oriented health systems. Thus, a proposed architecture can be treated as a decision support tool that directs health policy- and decision-makers toward more sustainable health systems. It can inform health policy makers and support them with formulating sustainable strategies at a national level. The agreed set of criteria can help to answer the question how well a health system is progressing towards social sustainability, and map its strengths and weaknesses in order to improve the performance.

The obvious popularity and simplicity of the AHP method in several decision problems and evaluation of various sustainability issues~\citep{dehe2015development} motivated the authors to extend the AHP method towards supporting the strong sustainability paradigm, namely Strong Sustainability Paradigm-AHP (SSP-AHP). According to the authors' best knowledge, this is the first attempt to limit the linear compensation of criteria in this very popular method. The SSP-AHP method proposed in the paper additionally provides a tool for modeling the strength of sustainability itself. As shown in the paper, SSP-AHP allows adjusting the influence of the decision-maker on the degree of reduction in criteria compensation depending on the problem that may require consideration of a more or less strong sustainability paradigm. The proposed method has great potential for use in any sustainability assessment problems in this context. Using the sustainability coefficient to model the strength of the reduction in the degree of criteria compensation is a useful tool that provides decision-makers with new analytical capabilities.

The authors also proved the usefulness of the SSP-AHP method in healthcare in practical terms. Based on the SSP-AHP method, a dedicated framework was developed to map sustainable value in the healthcare system. Its usefulness was demonstrated by evaluating selected European countries. Additionally, practical capabilities of the proposed/developed tool were presented, including modeling the strength of sustainability paradigm support. Besides structuring the model for Europe, the author's measure also includes the framework, where a set of criteria and weights is made available for a wider audience, which other researchers and decision-makers can use. 

The direction for further work is to implement the strong sustainability paradigm in the remaining MCDA methods. Additionally, the adaptation of fuzzy theory-based extensions of MCDA methods towards supporting a strong sustainability paradigm seems to be a promising research direction. From the methodical and practical point of view, the adaptation of the DEMATEL method to the proposed model and data structure and possible analysis of inter-criteria relationships can be a particularly interesting direction of further work. Adaptation of the hybrid method based on AHP and selected method from European school seems to be also interesting future work avenue. 

\appendix
\section{The Weighting Methods}
\label{sec:app}
\subsection{The AHP-based relative weighting method}
\label{sec:appWeights}
The AHP-based relative weighting method was detailed based on~\citep{ben2018risk}.
In the classical AHP method, the domain expert or decision-maker determines the criteria weights with a pairwise comparison of particular criteria regarding their respective significance.

\textbf{Step 1. } Decision-maker using the scale proposed by Saaty provided in Table~\ref{tab:saatyScale} expresses the degree of comparability between the two criteria.

\begin{table}[H]
\centering
\caption{The fundamental Saaty scale.}
\label{tab:saatyScale}
\resizebox{0.5\linewidth}{!}{
\begin{tabular}{ll} \toprule
Intensity of importance & Meaning \\ \midrule
1 & Equal importance \\
2 & Weak \\
3 & Moderate importance \\
4 & Moderate plus \\
5 & Strong importance \\
6 & Strong plus \\
7 & Very strong or demonstrated importance \\
8 & Very, very strong \\
9 & Extreme importance \\ \bottomrule
\end{tabular}
}
\end{table}

The results of the comparisons are collected in a pairwise comparison matrix $X = [x_{ij}]_{n \times n}$ presented in the Equation~(\ref{eq:matrixCriteriaComparision})

\begin{equation}
    X_{n \times n} = \left [ \begin{array}{cccccc} 
    1 & x_{12} & \cdots & \cdots & \cdots & x_{1n} \\
    \frac{1}{x_{12}} & 1 & \cdots & \cdots & \cdots & \cdots \\
    \cdots & \cdots & 1 & x_{ij} & \cdots & \cdots \\
    \cdots & \cdots & \frac{1}{x_{ij}} & 1 & \cdots & \cdots \\
    \cdots & \cdots & \cdots & \cdots & 1 & \cdots \\
    \frac{1}{x_{1n}} & \cdots & \cdots & \cdots & \cdots & 1 \\
    \end{array} \right ] \label{eq:matrixCriteriaComparision}
\end{equation}
where $x_{ij}$ denotes result of comparison between criterion $i$ and $j$ and obviously $x_{ij} = \frac{1}{x_{ji}}$ and $x_{ji} = \frac{1}{x_{ij}}$. Of course $x_{ii} = 1$ because each criterion is equally important to itself.

\textbf{Step 2.} Checking the consistency of the comparison matrix created in the previous step with index proposed by Saaty given in Equation~(\ref{eq:CI})

\begin{equation}
    CI(X) = \frac{\lambda_{max} - n}{n - 1} \label{eq:CI}
\end{equation}
where $\lambda_{max}$ denotes the maximum eigenvalue of the matrix presented in Equation~(\ref{eq:matrixCriteriaComparision}) and $n$ represents criteria number.

Then, the consistency ratio $CR$ is determined using Equation~(\ref{eq:CR})

\begin{equation}
    CR(X) = \frac{CI(X)}{RI_{n}} \label{eq:CR}
\end{equation}
and $RI_{n}$ means the random indices calculated and recommended by Saaty. Values of $RI_{n}$ are provided in Table~\ref{tab:RI}

\begin{table}[H]
\centering
\caption{Random indices $RI$ recommended by Saaty.}
\label{tab:RI}
\resizebox{0.5\linewidth}{!}{
\begin{tabular}{lrrrrrrrrrr} \toprule
n & 1 & 2 & 3 & 4 & 5 & 6 & 7 & 8 & 9 & 10 \\ \midrule
RI & 0 & 0 & 0.58 & 0.90 & 1.12 & 1.24 & 1.32 & 1.41 & 1.45 & 1.49 \\ \bottomrule
\end{tabular}
}
\end{table}

Following Saaty, matrices with $CR \leq 0.1$ are consistent and acceptable. On the other hand, $CR > 0.1$ indicates inconsistency of the considered matrix.

\textbf{Step 3. } Calculation of the priority vector of criteria. In this research, a vector was determined using the eigenvector method proposed by Saaty. A priority vector is the principal eigenvector of $X$ in this technique. The matrix $X$ contains ratios between weights and has to be multiplied by $w$, as Equation~\ref{eq:ahp3} presents.

\begin{equation}
    Xw = \left [ \begin{array}{cccc} 
    \frac{w_{1}}{w_{1}} & \frac{w_{1}}{w_{2}} & \cdots & \frac{w_{1}}{w_{n}} \\
    \frac{w_{2}}{w_{1}} & \frac{w_{2}}{w_{2}} & \cdots & \frac{w_{2}}{w_{n}} \\
    \vdots & \cdots & \cdots & \vdots \\
    \frac{w_{n}}{w_{1}} & \frac{w_{n}}{w_{2}} & \cdots & \frac{w_{n}}{w_{n}} \\
    \end{array} \right ] \left [ \begin{array}{c}
    w_{1} \\
    w_{2} \\
    \vdots  \\
    w_{n} \\
    \end{array} \right ] = \left [ \begin{array}{c}
    nw_{1} \\
    nw_{2} \\
    \vdots  \\
    nw_{n} \\
    \end{array} \right ] = nw \label{eq:ahp3}
\end{equation}
Equation $Xw = nw$ implies that $n$ and $w$ denote an eigenvalue and an eigenvector of $X$ and $n$ is the largest eigenvalue of $X$. Vector $w$ can be received from a pairwise comparison matrix $X$ by solving the below equation system, as Equation~\ref{eq:eigenvec} demonstrates.

\begin{equation}
    \left \{ \begin{array}{c}
    Xw = \lambda_{max}\\
    w^{T} = [1, 1, \ldots, 1]^{T} = 1
    \end{array} \right. \label{eq:eigenvec}
\end{equation}

The principal eigenvector of $X$ can be calculated simply using a Python software package provided in codes available in supplementary materials at GitHub~\citep{sspahpgithub2022}.

\subsection{The Entropy weighting method}

The subsequent steps of Entropy weighting method are provided below, based on~\citep{lotfi2010imprecise}.

\textbf{Step 1. } Normalization of the decision matrix $X = [x_{ij}]_{m \times n}$

\begin{equation}
    p_{ij} = \frac{x_{ij}}{\sum_{i=1}^{m}x_{ij}} \label{eq:entropyNormalization}
\end{equation}

\textbf{Step 2. } Calculation of the entropy $h_{j}$

\begin{equation}
    h_{j} = -(ln \; m)^{-1}\sum_{i=1}^{m}p_{ij}\cdot ln \; p_{ij}
\end{equation}
where $p_{ij}\cdot ln \; p_{ij}$ is established as 0 if $p_{ij} = 0$

\textbf{Step 3. } Calculation of weights values

\begin{equation}
    w_{j} = \frac{1 - h_{j}}{\sum_{j=1}^{n}(1 - h_{j})}
\end{equation}

\subsection{The CRITIC weighting method}

The subsequent stages of Criteria Importance Through Inter criteria
Correlation weighting method (CRITIC) are in following steps based on~\citep{tucs2019new}.
Decision matrix $X = {[x_{ij}]}_{m \times n}$ contains $m$ alternatives and $n$ criteria, where $x_{ij}$ represents the performance measure of $i^{th}$ alternative with regard to $j^{th}$ criterion. To determine the weight of the $j^{th}$ criterion $w_{j}$ using CRITIC the calculations provided below are conducted

\noindent \textbf{Step 1.} Decision matrix is normalized by Equation~(\ref{eq:critic1a}) as for benefit criteria

\begin{equation}
    r_{ij} = \frac{x_{ij}-min_{j}(x_{ij})}{max_{j}(x_{ij})-min_{j}(x_{ij})} \label{eq:critic1a}
\end{equation}

\noindent \textbf{Step 2.} Calculation of the Pearson correlation coefficient among pairs of criteria as Equation~(\ref{eq:critic2}) demonstrates.

\begin{equation}
    \rho _{jk} = \frac{\sum_{i=1}^{m}(r_{ij} - \overline{r}_{j})(r_{ik} - \overline{r}_{k})}{\sqrt{\sum_{i=1}^{m}(r_{ij} - \overline{r}_{j})^{2} \sum_{i=1}^{m}(r_{ik} - \overline{r}_{k})^{2}}}. \label{eq:critic2}
\end{equation}

\noindent \textbf{Step 3.} Calculation of criteria weights using Equation~(\ref{eq:critic3}) and~(\ref{eq:critic4}),

\begin{equation}
    c_{j} = \sigma _{j} \sum_{k=1}^{n} (1 - \rho _{jk}); \label{eq:critic3}
\end{equation}

\begin{equation}
    w_{j} = \frac{c_{j}}{\sum_{k=1}^{n}c_{k}}, \label{eq:critic4}
\end{equation}
\noindent
where $i = 1, 2, \ldots, m; \;  j, k = 1, 2, \ldots, n.$
In the formulas given above $c_{j}$ represents the quantity of information contained in $j^{th}$ criterion, $\sigma_{j}$ express the standard deviation of the $j^{th}$ criterion and $\rho_{jk}$ is the correlation coefficient between the $j^{th}$ and $k^{th}$ criteria. A high standard deviation and low correlation of given criterion with the others determine a high criterion weight. Thus, a high value of $C_j$ provides more information from the considered criterion

\section{Characteristics of the healthcare experts involved in the research}
\label{app:experts}

A total of 30 health care experts were invited to the study. The recruitment of experts was preceded by the creation of address database, which included health sector professionals with experience in assessing health systems. Social media (like Linkedin) have been used to contact potential respondents. Finally 16 experts agreed to take part in on-line survey in October 2021. The most represented professional group was the management of medical institutions (5 people) and public administration staff (4 people). The vast majority of the invited health sector professionals held doctoral degrees (8 people). They were experts with extensive professional experience in the health sector, most of them more than 20 years. Among the participants of the study, men and those working in the public sector predominated. The sample structure is presented in Table~\ref{tab:experts} provided below. 

\begin{table}[H]
\centering
\caption{Demographic characteristics  of the sample group (N)}
\label{tab:experts}
\resizebox{0.6\linewidth}{!}{
\begin{tabular}{llr} \toprule
\multirow{2}{*}{Gender} & Male & 9 \\
 & Female & 7 \\ \hline
\multirow{3}{*}{Age} & <40 & 2 \\
 & 40-50 & 6 \\
 & >50 & 8 \\ \hline
\multirow{6}{*}{Dominant background} & Healthcare entities   -  managerial position & 5 \\
 & Healthcare sector-  administrative position & 4 \\
 & Healthcare unit-   medical staff & 2 \\
 & Insurance scheme & 1 \\
 & Government – department of healthcare; ministry & 3 \\
 & Third party payer & 2 \\ \hline
\multirow{4}{*}{Education} & MBA & 5 \\
 & Ph.D. & 8 \\
 & Professor & 1 \\
 & M.Sc. & 2 \\ \hline
\multirow{3}{*}{Workplace} & Public sector   employee & 13 \\
 & Private sector   employee & 2 \\
 & Entrepreneur, self-employed & 1 \\ \hline
\multirow{3}{*}{Years of   experience} & 10-15 & 2 \\
 & 16-20 & 5 \\
 & > 20 & 9 \\ \hline
\multirow{11}{*}{Country of   origin} & Poland & 2 \\
 & Hungary & 2 \\
 & Latvia & 2 \\
 & France & 1 \\
 & Belgium & 1 \\
 & Finland & 1 \\
 & Sweden & 1 \\
 & Norway & 1 \\
 & Slovenia & 2 \\
 & United Kingdom & 2 \\
 & Germany & 1 \\ \bottomrule
\end{tabular}
}
\end{table}

\section{Criteria weights}
\label{app:resultsWeights}
\begin{table}[H]
\centering
\caption{Criteria weights determined using different weighting methods.}
\label{tab:resultsWeights}
\resizebox{0.3\linewidth}{!}{
\begin{tabular}{lrrr} \toprule
$C_{i}$ & AHP-based relative & CRITIC & Entropy \\ \midrule
$C_{1}$ & 0.0615 & 0.0330 & 0.0253 \\
$C_{2}$ & 0.0615 & 0.0317 & 0.0170 \\
$C_{3}$ & 0.0615 & 0.0328 & 0.0100 \\
$C_{4}$ & 0.0615 & 0.0330 & 0.0408 \\
$C_{5}$ & 0.0410 & 0.0476 & 0.0300 \\
$C_{6}$ & 0.0410 & 0.0355 & 0.0596 \\
$C_{7}$ & 0.0410 & 0.0289 & 0.0671 \\
$C_{8}$ & 0.0296 & 0.0482 & 0.0762 \\
$C_{9}$ & 0.0296 & 0.0407 & 0.0010 \\
$C_{10}$ & 0.0296 & 0.0410 & 0.0124 \\
$C_{11}$ & 0.0296 & 0.0500 & 0.0356 \\
$C_{12}$ & 0.1182 & 0.0381 & 0.0347 \\
$C_{13}$ & 0.0591 & 0.0409 & 0.0003 \\
$C_{14}$ & 0.0591 & 0.0567 & 0.0508 \\
$C_{15}$ & 0.0646 & 0.0401 & 0.1064 \\
$C_{16}$ & 0.0323 & 0.0464 & 0.0496 \\
$C_{17}$ & 0.0323 & 0.0415 & 0.0694 \\
$C_{18}$ & 0.0595 & 0.0471 & 0.0003 \\
$C_{19}$ & 0.0298 & 0.0377 & 0.0024 \\
$C_{20}$ & 0.0298 & 0.0380 & 0.0550 \\
$C_{21}$ & 0.0047 & 0.0411 & 0.0237 \\
$C_{22}$ & 0.0047 & 0.0368 & 0.0869 \\
$C_{23}$ & 0.0047 & 0.0357 & 0.0570 \\
$C_{24}$ & 0.0047 & 0.0420 & 0.0728 \\
$C_{25}$ & 0.0094 & 0.0353 & 0.0156 \\ \bottomrule
\end{tabular}
}
\end{table}

\section{The Correlation Coefficients}
\label{sec:appCorrelations}
\subsection{Weighted Spearman's Rank Correlation Coefficient}

The calculation of the weighted Spearman correlation coefficient is shown by the Equation~\ref{wSpearCoeff}, where $N$ is a sample size, $x_i$, and $y_i$ are the values in the compared rankings~\citep{salabun2020mcda}. 
\begin{equation}
    r_{w} = 1 - \frac{6\sum_{i=1}^{N}(x_{i} - y_{i})^{2} ((N - x_{i} + 1) + (N - y_{i} + 1))}{N^{4} + N^{3} - N^{2} - N} \label{wSpearCoeff}
\end{equation}

\subsection{The Pearson Correlation Coefficient}

The Pearson correlation coefficient is represented by the Equation~\ref{pearsonCoeff}
\begin{equation}
    r_{xy} = \frac{N \sum x_{i}y_{i} - \sum x_{i}\sum y_{i}}{\sqrt{N\sum x_{i}^{2} - (\sum x_{i})^{2}}\sqrt{N\sum y_{i}^{2}-(\sum y_{i})^{2}}} \label{pearsonCoeff}
\end{equation}
where $N$ is a sample size, $x_i$ and $y_i$ are the individual sample elements indexed with $i$~\citep{deng2021combining}.

\section*{Acknowledgments}
This research was funded in part by National Science Centre, Poland 2022/45/B/HS4/02960. For the purpose of Open Access, the author has applied a CC-BY public copyright licence to any Author Accepted Manuscript (AAM) version arising from this submission


\bibliography{mybibfile-clean}








\end{document}